\documentclass[manuscript]{emulateapj}

\usepackage{aas_macros}
\usepackage{natbib}
\usepackage{url}

\usepackage{multirow}
\usepackage{graphicx}
\usepackage{amsmath, amsthm, amssymb}
\usepackage{color}
\usepackage{float}
\usepackage{enumerate}
\usepackage{textcomp}

\shorttitle{Magnetized GMC Collisions}
\shortauthors{Wu et al.}

\begin{document}

\title{GMC Collisions as Triggers of Star Formation. I.\\ Parameter space exploration with 2D simulations} 

\author{Benjamin Wu} 
\affil{Department of Physics,
        University of Florida,
        Gainesville, FL 32611, USA}
\email{benwu@phys.ufl.edu}

\author{Sven Van Loo} 
\affil{School of Physics and Astronomy, University of Leeds, 
        Leeds LS2 9JT, UK}
\affil{Harvard-Smithsonian Center for Astrophysics, 60 Garden St.,
        Cambridge, MA 02138, USA}

\author{Jonathan C. Tan} 
\affil{Departments of Astronomy \& Physics,
        University of Florida,
        Gainesville, FL 32611, USA}

\and

\author{Simon Bruderer} 
\affil{Max Planck Institute for Extraterrestrial Physics,
        Giessenbachstrasse 1, D-85748 Garching, Germany}

\begin{abstract}
We utilize magnetohydrodynamic (MHD) simulations to develop a
numerical model for GMC-GMC collisions between nearly magnetically
critical clouds.  The goal is to determine if, and under what
circumstances, cloud collisions can cause pre-existing magnetically
subcritical clumps to become supercritical and undergo gravitational
collapse. 
We first develop and implement new photodissociation region (PDR)
based heating and cooling functions that span the atomic to molecular
transition, creating a multiphase ISM and allowing modeling of
non-equilibrium temperature structures.
Then in 2D and with ideal MHD, we explore a wide parameter space of
magnetic field strength, magnetic field geometry, collision velocity,
and impact parameter, and compare isolated versus colliding clouds.
We find factors of $\sim 2-3$ increase in mean clump density from
typical collisions, with strong dependence on collision velocity and
magnetic field strength, but ultimately limited by flux-freezing in 2D
geometries. For geometries enabling flow along magnetic field lines,
greater degrees of collapse are seen.
We discuss observational diagnostics of cloud collisions, focussing on
$^{13}$CO($J$=2-1), $^{13}$CO($J$=3-2), and $^{12}$CO($J$=8-7) integrated 
intensity maps and spectra, which we synthesize from our simulation 
outputs. We find the ratio of $J$=8-7 to lower-$J$ emission is a powerful
diagnostic probe of GMC collisions.
\end{abstract}


\section{Introduction}
\label{sec:intro}

Understanding star formation is a key astrophysical problem, with many
fundamental questions still unresolved.  In particular, what
mechanisms drive or inhibit the star formation process?  Studying the
evolution of giant molecular clouds (GMCs) within the diffuse
interstellar medium and the formation of prestellar clumps and cores
within GMCs is complicated because of the large range of densities
($n_{\rm H} \sim 1\:{\rm cm^{-3}}$ to $\sim 10^{6}\:{\rm cm^{-3}}$),
length scales ($\sim$kpc to $\sim$0.1~pc), and timescales ($\sim
10^{8}$~yr for galactic orbits to $\sim 10^{5}$~yr for core dynamical
timescales) involved, as well as the nonlinear effects of
self-gravity, thermal, magnetic, and turbulent pressures, large-scale
motions such as galactic shear or collisional converging flows,
radiation, chemistry, and feedback. Additionally, the initial
conditions are uncertain and boundary conditions are poorly
constrained. The final conditions, gleaned through observations,
suggest that star formation is highly clustered and localized, with
relatively high local efficiency within clusters
\citep[e.g.,][]{Lada_Lada_2003,Myers_Allen_Pipher_Fazio_2009}. However,
overall star formation is slow and inefficient, with only a few
percent of total gas forming stars over local dynamical timescales
\citep[e.g.,][]{Zuckerman_Evans_1974,Krumholz_Tan_2007,DaRio_ea_2014}.

With regard to the importance of magnetic fields \citep[see,
  e.g.,][]{Crutcher_2012,Li_ea_2014}, there have been two main views.
Strong-field models propose relatively long GMC lifetimes in which
magnetic fields play important roles in controlling formation and
evolution of the clouds.
In these models, non-star-forming clouds are initially subcritical,
i.e., their magnetic fields are strong enough to prevent gravitational
collapse.  Weak-field models have GMCs as intermittent phenomena with
short lifetimes ($\sim10^{6}$~yr) and turbulent flows controlling the
formation of clouds, clumps and cores.  These models posit
magnetically supercritical masses, i.e., the magnetic pressure alone
is too weak to support against gravity.

Zeeman measurements show that mass-to-magnetic flux ratios ($M/\Phi$)
are approximately critical to slightly supercritical in molecular
clouds
\citep{Crutcher_1999,Troland_Crutcher_2008,Crutcher_2012,Li_ea_2014}.
If GMCs are partially stabilized by magnetic fields, then this may
increase their lifetimes to $\gtrsim 20$~Myr timescales, which are
then comparable to GMC-GMC collision times, especially for clouds
inside the solar circle
\citep{Gammie_ea_1991,Tan_2000,Tasker_Tan_2009,Dobbs_ea_2015}.
Indirect observational evidence for frequent GMC collisions comes from
the near random orientations of projected angular momentum vectors of
GMCs
\citep{Rosolowsky_ea_2003,Koda_ea_2006,Imara_Blitz_2011,Imara_Bigiel_Blitz_2011}.

Frequent GMC collisions could be an important mechanism for injecting
turbulent energy into GMCs \citep{Tan_2000,Tan_Shaske_Van_Loo_2013},
with the energy being extracted from galactic orbital motion. Other
mechanisms for injecting turbulence involve star formation feedback
\citep{Matzner_2007,Goldbaum_ea_2011}.  Without such replenishment,
turbulence is expected to decay within about a crossing time
\citep{Low_Smith_Klessen_Burkert_1998,Ostriker_Gammie_Stone_1999}.

By producing dense gas, compressed in shocks, GMC-GMC collisions may
be an important trigger of star cluster formation
\citep{Scoville_Sanders_Clemens_1986}. If the majority of star
formation is initiated by this process, then a model of shear-mediated
cloud collisions can naturally explain the observed dynamical
Kennicutt-Schmidt relation \citep{Kennicutt_1998} in which a roughly
constant fraction of gas, $\epsilon_{\rm orb}\simeq0.04$ is converted
into stars every local orbital time 
\citep{Tan_2000,Tan_2010,Suwannajak_ea_2014}. 
Note that this mechanism of creating star-forming molecular
clumps from localized compressed regions of pre-existing GMCs, is
different from that proposed for creating molecular clouds from shocks
in converging flows of atomic gas
\citep[e.g.,][]{Heitsch_ea_2006,Loo_Falle_Hartquist_Moore_2007,Heitsch_Stone_Hartmann_2009,Van_Loo_Falle_Hartquist_2010}.

Cloud-cloud collisions have been investigated by a number of previous
studies. \citet{Habe_Ohta_1992} performed 2D axisymmetric SPH
simulations of head-on collisions of non-identical clouds. These
collisions produced a bow shock which disrupted the larger cloud while
compressing the smaller cloud. This compression could lead to
gravitational instability for the smaller cloud even if its initial
mass was below the Jeans mass.

\citet{Klein_Woods_1998} presented 2D AMR hydrodynamics simulations of
homogeneous cloud collisions. The collisions resulted in bending mode
instabilities creating large aspect ratio filaments. With surface
perturbations, the merged cloud system became highly asymmetrical and
highly inhomogeneous with islands of high density surrounded by low
density regions.

\citet{Anathpindika_2009} performed a series of 3D SPH simulations which 
investigated the gravitational stability of post-shock compressed slabs
resulting from molecular cloud collisions. Additionally, sheared collisions
result in non linear thin shell instabilities and Kelvin-Helmholtz 
instabilities.

More recently, \citet{Takahira_ea_2014} performed 3D hydrodynamic
simulations with AMR showing core formation occurring from a GMC
collision interface. They found that faster collision velocities
formed a greater number of cores, but core growth was predominantly
via accretion in the shock front, with slower shocks being favored for
making larger cores.

In terms of MHD collision studies, \citet{Koertgen_Banerjee_2015}
investigated molecular cloud formation and the transition from magnetically 
sub- to supercritical HI clouds via converging magnetized flows.
Even with magnetic diffusion effects, they found that cylindrical flows
created no magnetically supercritical regions and star formation is strongly
suppressed for even relatively low initial magnetic field strengths.

This paper, the first of a series, explores the process of magnetized
cloud-cloud collisions and its effect on individual GMC and clump
scales.  Here we restrict analysis to a parameter space exploration
with 2D simulations of simplified cloud geometries, including an
embedded clump: formally colliding infinite cylinders, which can
approximate collisions of spheroidal clouds.

\S\ref{sec:methods} explains the fiducial set-up and various
simulation and analytic methods employed.  \S\ref{sec:heatcool}
describes new heating and cooling functions that we have developed for
this project.  \S\ref{sec:results} describes the set up and subsequent
results of exploring the following parameters: magnetic field
strength, magnetic field orientation, collision velocity,
and impact parameter. \S\ref{sec:diagnostics} discusses predictions of
observational diagnostics of shocks.  Discussion and conclusions
follow in \S\ref{sec:conclusion}.

\section{Numerical Model}
\label{sec:methods}

\subsection{Initial Conditions}

For our default initial conditions we use typical observed values of
Galactic GMC and ISM properties. GMCs are conventionally defined as having
masses $\geq10^4\:M_\odot$. They have mean mass surface densities
$\Sigma \sim 100\:M_{\odot}\:{\rm pc^{-2}}$
\citep[e.g.,][]{McKee_Ostriker_2007,Tan_Shaske_Van_Loo_2013}. Typical
mean volume densities are $n_{\rm H} \simeq 100\:{\rm cm^{-3}}$,
although clumps and cores within the clouds have densities that can be
orders of magnitude larger. 

GMCs have internal velocity dispersions that are similar to virial
velocities, typically several km/s, which is much larger than the
$\sim 0.2$~km/s sound speeds of $\sim 10$~K gas. Supersonic turbulence
and self-gravity are thought to be two important processes that help
give rise to the hierarchical density structures seen in
GMCs. However, these structures may also be regulated by magnetic
fields.

The magnetic field of the local diffuse ISM background is $6\pm2~{\rm
  \mu G}$ \citep{Beck_2001}. If a random, uniform distribution of
field strengths is assumed up to a maximum value, $B_{\rm max}$,
Zeeman measurements reveal that this maximum magnetic field value
measured within molecular clouds, clumps and cores with $n_{\rm
  H}>300\:{\rm cm^{-3}}$ scales as $B_{\rm max} = B_{0}(n_{\rm
  H}/300\:{\rm cm^{-3}})^{0.65}$, where $B_{0} = 10~{\rm \mu G}$
\citep{Crutcher_ea_2010}. At lower densities, $B_{\rm max} = B_{0} =
10~{\rm \mu G}$, independent of density. We will henceforth refer to
this as the ``Crutcher relation''. Thus, for $n_{\rm H} = 10^3~{\rm
  cm^{-3}}$, $B_{\rm max} \simeq 22~{\rm \mu G}$.

Observed random velocities of Galactic GMCs are approximately
5-7~${\rm km\ s^{-1}}$
\citep[e.g.,][]{Liszt_ea_1984,Stark_1984}. However, interaction
velocities between colliding GMCs are likely to be set by the shear
velocity at 1 to 2 tidal radii of the clouds \citep{Gammie_ea_1991,Tan_2000},
which can be several times larger.

Given the 2D nature of the simulations of this paper, the modeled
structures can be considered as filaments extending perpendicular to
the simulation domain. We follow ``clouds'', i.e., ``GMCs,'' with a
uniform density of $n_{\rm H,GMC} = 100\:{\rm cm^{-3}}$.  Although the
clouds are, in principle, cylinders of infinite extent, we assume that
the clouds have a finite mass, i.e., $M_{\rm GMC} = 10^5\:M_\odot$. A
mass surface density of $\Sigma = 100\:M_\odot\:{\rm pc^{-2}}$
integrated along the cloud axis, then, gives a typical cloud radius
$R_{\rm GMC} = 17.8$~pc.  
We set the radius of the first cloud, i.e.,
Cloud~1, to $R_{1} = 23.8\:{\rm pc}$ and of the second one, i.e.,
Cloud~2, to $R_{2} = 0.5 R_1 = 11.9\:{\rm pc}$.

\begin{figure}[h]
\includegraphics[width=1.0\columnwidth]{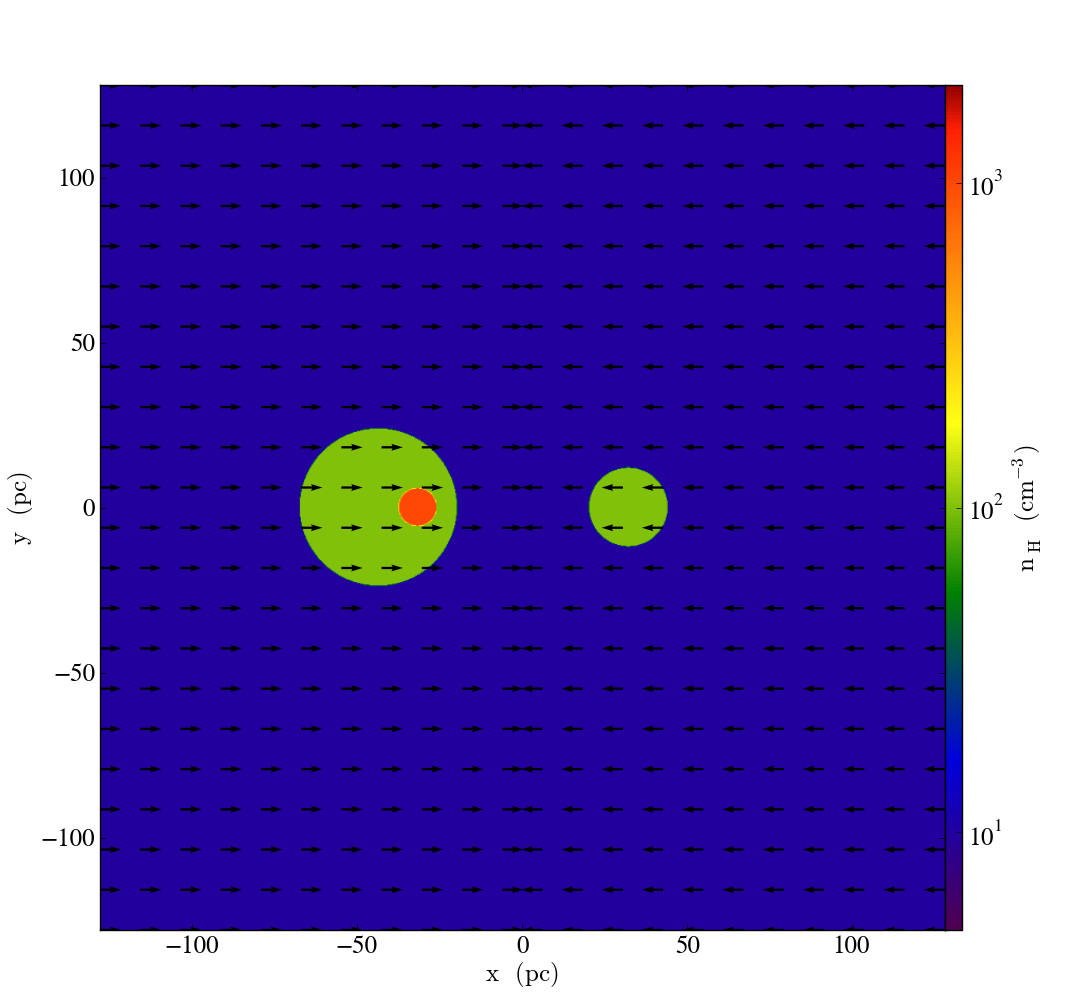}
\begin{center}
\caption{\label{fig:initial}
Basic cloud collision setup. GMC 1 (left cloud) has radius $R_1$.  
It includes an embedded clump with radius $R_{\rm cl}$ located at a 
distance of one half-radius to the edge.  GMC 1 collides with GMC 2,
a uniform cloud with a radius $R_2$ = $R_1$/2.  
The clouds are initially separated 
by a distance that is changed depending on
relative velocity so that collisions occur at the same time.}
\end{center}
\end{figure}

GMCs are structured, containing dense clumps. When a collision occurs,
the effect of this collision on pre-existing clumps may be the most
important for triggering star formation. We therefore introduce an
idealized embedded, overdense clump into Cloud~1. The clump has a
uniform density of $n_{\rm H,cl} = 1000~{\rm cm^{-3}}$, i.e., $10\times$
overdense compared to the GMC, and a radius of 5.6~pc. We position the
clump off center at $(x,y) = (0.5R_1, 0)$ (see
Fig.~\ref{fig:initial}). The properties of our clouds and clump are
listed in Table~\ref{tab:GMC-properties}.

\begin{table}
\caption{GMC and Clump Properties}
\label{tab:GMC-properties}
\begin{center}
\begin{tabular}{lcccc}
\hline\hline
 & $M_{\rm tot}$\footnote{Masses are for an equivalent spherical cloud.} & $n_{\rm H}$ & $R$ & $B_{\rm crit}$\footnote{Critical $B$-field strengths are listed for the GMCs and the clump, along with the fiducial ambient field strength.}\\
 & $(10^{5}\:M_{\odot})$  & $({\rm cm}^{-3})$  & ${\rm (pc)}$ & ${(\rm \mu G)}$ \\
\hline
Ambient & - & 10 & - & 10\\
GMC 1\footnote{GMC 1 includes a clump, but properties listed here are for non-clump material within the cloud} & 1.78 & 100  & 23.8 & 42.0\\
GMC 2 & 0.56 & 100 & 11.9 & 13.9\\ 
Clump & 0.10 & 1000 & 5.64 & 66.0\\
\hline
\end{tabular}
\end{center}
\end{table}

For typical molecular cloud temperatures, $\sim 10-20$~K, such GMCs
and clumps are not thermally supported against gravitational collapse.
For example, if the clouds are considered as long filaments in the
direction perpendicular to the simulation plane, then the mass per
unit length for 
GMC 1, $m_{l} = 6150\:M_\odot\:{\rm pc^{-1}}$, far exceeds the critical
line mass for a cylindrical cloud, given by $m_{l,{\rm crit}}
= 2c_s^2/G$, which is $\sim20\:M_\odot\:{\rm pc}^{-1}$ for a cold $10$~K
cloud \citep{Ostriker_1964}.  

The inclusion of magnetic fields helps to stabilize the clouds.  We
vary the direction, i.e., parallel, perpendicular and oblique to the
cylindrical axis of the cloud, and the magnitude of the field. The
field strengths are detailed in the results section, but are of the
order of $10-100\:{\rm \mu G}$, given observed field strengths
\citep{Crutcher_2012}.

Internal cloud turbulence is another mechanism that may help support
GMCs. To separate out the effects of magnetic fields from turbulence,
in this paper we do not initialize the clouds with any turbulence,
deferring this to Paper II, which also extends the dimensionality to
3D. However, turbulent motions are generated by the GMC-GMC collision,
which may then provide additional support to the clouds.

The ambient medium in which the clouds are embedded, representative of
the atomic cold neutral medium, is set to have $n_{\rm H,0} = 10\:{\rm
  cm^{-3}}$, with a magnetic field of $B_{0}=10\:{\rm \mu G}$.
The default relative collision velocity of the clouds is set to be
$v_{\rm rel}=10\:{\rm km\:s^{-1}}$, with variations from 5 to 40~${\rm
  km\ s^{-1}}$.
The default impact parameter, $b$, of the collision is set to zero,
i.e., an on-axis collision, but some cases with $b=0.5, 1, 1.5\:R_1$
are also explored.
The surrounding medium, which we consider to be a co-moving atomic
envelope around the GMC, is also colliding. Thus in terms of the
simulation domain, half the box is moving with $+v_{\rm rel }/2$ and
the other half has $-v_{\rm rel}/2$.
However, at faster velocities $\gtrsim 20\:{\rm km\:s^{-1}}$ we
sometimes notice modest effects of numerical viscosity on clump
properties and so also run simulations in the velocity frame of Cloud
1.

A summary of key parameters in all runs performed in this paper is
listed in Table~\ref{tab:all_runs}.
Velocities denoted with a ``*'' indicate models run in the frame of
Cloud 1.

\begin{table*}[h]
\caption{Summary of Simulation Run Parameters}
\label{tab:all_runs}
\begin{center}
\begin{tabular}{lcccccccccc} \hline \hline
Run & $n_{\rm H,0}$ & $n_{\rm H,1}$ & $n_{\rm H,cl}$ & $n_{\rm H,2}$ & 
$\mathbf{B_{\rm 0}}$ & $\mathbf{B_{\rm 1}}$  & $\mathbf{B_{\rm cl}}$ & $\mathbf{B_{\rm 2}}$ & $v_{\rm rel}$ & $b$\\
 & (cm$^{-3}$) & (cm$^{-3}$) & (cm$^{-3}$) & (cm$^{-3}$) & (${\rm \mu G}$) & (${\rm \mu G}$) & (${\rm \mu G}$) & (${\rm \mu G}$) & (km/s) & ($R_{1}$)\\ 
\hline \hline
\multicolumn{11}{c}{0. Out-of-plane Fields $(0,0,B_{z})$} \\ 
\hline 
\multicolumn{11}{l}{0. Isolated Cloud} \\
    0.A.0     & 10    & 100   & -     & -     & (0,0,10)     & (0,0,27.9)   & -     & -     & -     & -       \\
    0.A.1     & 10    & 100   & -     & -     & (0,0,0)      & (0,0,0)      & -     & -     & -     & -       \\
    0.A.2     & 10    & 100   & -     & -     & (0,0,10)     & (0,0,10)     & -     & -     & -     & -       \\
    0.A.3     & 10    & 100   & -     & -     & (0,0,10)     & (0,0,20)     & -     & -     & -     & -       \\
    0.A.4     & 10    & 100   & -     & -     & (0,0,10)     & (0,0,40)     & -     & -     & -     & -       \\
    \noalign{\smallskip}
    \hline
    \multicolumn{11}{c}{1. Out-of-plane Fields $(0,0,B_{z})$} \\ 
    \hline
    \multicolumn{11}{l}{1.B. Isolated Cloud with Clump} \\
    1.B.0     & 10    & 100   & 1000  & -     & (0,0,10)     & (0,0,40)     & (0,0,65)     & -     & -     & -       \\
    1.B.1.1   & 10    & 100   & 1000  & -     & (0,0,10)     & (0,0,10)     & (0,0,65)     & -     & -     & -       \\
    1.B.1.2   & 10    & 100   & 1000  & -     & (0,0,10)     & (0,0,20)     & (0,0,65)     & -     & -     & -       \\
    1.B.1.3   & 10    & 100   & 1000  & -     & (0,0,10)     & (0,0,65)     & (0,0,65)     & -     & -     & -       \\
    1.B.2.1   & 10    & 100   & 1000  & -     & (0,0,10)     & (0,0,40)     & (0,0,40)     & -     & -     & -       \\
    1.B.2.2   & 10    & 100   & 1000  & -     & (0,0,10)     & (0,0,40)     & (0,0,55)     & -     & -     & -       \\
    1.B.2.3   & 10    & 100   & 1000  & -     & (0,0,10)     & (0,0,40)     & (0,0,75)     & -     & -     & -       \\
    \noalign{\smallskip}
    \multicolumn{11}{l}{1.C. Cloud Collision} \\
    1.C.0     & 10    & 100   & 1000  & 100   & (0,0,10)   & (0,0,40)   & (0,0,65)   & (0,0,13.2)   & 10    & 0      \\
    1.C.1.1   & 10    & 100   & 1000  & 100   & (0,0,10)   & (0,0,40)   & (0,0,65)   & (0,0,13.2)   & 5     & 0      \\
    1.C.1.2   & 10    & 100   & 1000  & 100   & (0,0,10)   & (0,0,40)   & (0,0,65)   & (0,0,13.2)   & 20*    & 0      \\
    1.C.1.3   & 10    & 100   & 1000  & 100   & (0,0,10)   & (0,0,40)   & (0,0,65)   & (0,0,13.2)   & 40*    & 0      \\
    1.C.2.2   & 10    & 100   & 1000  & 100   & (0,0,10)   & (0,0,40)   & (0,0,65)   & (0,0,13.2)   & 10    & 0.5    \\
    1.C.2.4   & 10    & 100   & 1000  & 100   & (0,0,10)   & (0,0,40)   & (0,0,65)   & (0,0,13.2)   & 10    & 1      \\
    1.C.2.5   & 10    & 100   & 1000  & 100   & (0,0,10)   & (0,0,40)   & (0,0,65)   & (0,0,13.2)   & 10    & 1.5    \\
    \noalign{\smallskip}
    \hline
    \multicolumn{11}{c}{2. In-plane Fields $(B_{x},0,0)$ and $(0,B_{y},0)$} \\
    \hline
    \multicolumn{11}{l}{2.B. Isolated Cloud with Clump} \\
    2.B.1.0   & 10    & 100   & 1000  & -     & (40,0,0)    & (40,0,0)   & (40,0,0)      & -     & -     & -       \\
    2.B.1.1   & 10    & 100   & 1000  & -     & (10,0,0)    & (10,0,0)   & (10,0,0)      & -     & -     & -       \\
    2.B.1.2   & 10    & 100   & 1000  & -     & (65,0,0)    & (65,0,0)   & (65,0,0)      & -     & -     & -       \\
    2.B.2.0   & 10    & 100   & 1000  & -     & (0,40,0)    & (0,40,0)   & (0,40,0)      & -     & -     & -       \\
    2.B.2.1   & 10    & 100   & 1000  & -     & (0,10,0)    & (0,10,0)   & (0,10,0)      & -     & -     & -       \\
    2.B.2.2   & 10    & 100   & 1000  & -     & (0,65,0)    & (0,65,0)   & (0,65,0)      & -     & -     & -       \\
    \noalign{\smallskip}
    \multicolumn{11}{l}{2.C. Cloud Collision} \\
    2.C.1.0   & 10    & 100   & 1000  & 100   & (40,0,0)   & (40,0,0)   & (40,0,0)   & (40,0,0)   & 10    & 0       \\
    2.C.1.1   & 10    & 100   & 1000  & 100   & (10,0,0)   & (10,0,0)   & (10,0,0)   & (10,0,0)   & 10    & 0       \\
    2.C.1.2   & 10    & 100   & 1000  & 100   & (65,0,0)   & (65,0,0)   & (65,0,0)   & (65,0,0)   & 10    & 0       \\
    2.C.2.0   & 10    & 100   & 1000  & 100   & (0,40,0)   & (0,40,0)   & (0,40,0)   & (0,40,0)   & 10    & 0       \\
    2.C.2.1   & 10    & 100   & 1000  & 100   & (0,10,0)   & (0,10,0)   & (0,10,0)   & (0,10,0)   & 10    & 0       \\
    2.C.2.2   & 10    & 100   & 1000  & 100   & (0,35,0)   & (0,65,0)   & (0,65,0)   & (0,65,0)   & 10    & 0       \\
    \noalign{\smallskip}
    \hline
    \multicolumn{11}{c}{3. Mixed Fields $(B_{x},B_{y},B_{z})$} \\
    \hline
    \multicolumn{11}{l}{3.B. Isolated Cloud with Clump} \\
    3.B.1     & 10    & 100   & 1000  & -     & (10,0,0)   & (10,0,38.7)   & (10,0,64.2)   & -     & -     & -     \\
    3.B.2     & 10    & 100   & 1000  & -     & (0,10,0)   & (0,10,38.7)   & (0,10,64.2)   & -     & -     & -     \\
    \noalign{\smallskip}
    \multicolumn{11}{l}{3.C. Cloud Collision} \\
    3.C.1.0   & 10    & 100   & 1000  & 100   & (10,0,0)   & (10,0,38.7)  & (10,0,64.2)  & (10,0,12.9)  & 10  & 0   \\
    3.C.1.1   & 10    & 100   & 1000  & 100   & (10,0,0)   & (10,0,38.7)  & (10,0,64.2)  & (10,0,12.9)  & 5   & 0   \\
    3.C.1.2   & 10    & 100   & 1000  & 100   & (10,0,0)   & (10,0,38.7)  & (10,0,64.2)  & (10,0,12.9)  & 20*  & 0   \\
    3.C.1.3   & 10    & 100   & 1000  & 100   & (10,0,0)   & (10,0,38.7)  & (10,0,64.2)  & (10,0,12.9)  & 40*  & 0   \\
    3.C.2.0   & 10    & 100   & 1000  & 100   & (0,10,0)   & (0,10,38.7)  & (0,10,64.2)  & (0,10,12.9)  & 10  & 0   \\
    3.C.2.1   & 10    & 100   & 1000  & 100   & (0,10,0)   & (0,10,38.7)  & (0,10,64.2)  & (0,10,12.9)  & 5   & 0   \\
    3.C.2.2   & 10    & 100   & 1000  & 100   & (0,10,0)   & (0,10,38.7)  & (0,10,64.2)  & (0,10,12.9)  & 20*  & 0   \\
    3.C.2.3   & 10    & 100   & 1000  & 100   & (0,10,0)   & (0,10,38.7)  & (0,10,64.2)  & (0,10,12.9)  & 40*  & 0   \\
    \noalign{\smallskip}
    \multicolumn{11}{l}{3.D. Cloud Collision with Impact Parameter} \\
    3.D.0     & 10    & 100   & 1000  & 100   & (10,0,0)   & (10,0,38.7)    & (10,0,64.2)    & (10,0,12.9)  & 10  & 0.5 \\
        \end{tabular}
    \end{center}
\end{table*}

\subsection{Numerical Code}\label{sec:num_method}

The numerical code is a modified version of the Adaptive Mesh
Refinement (AMR) code Enzo 2.0
\citep{BryanNorman1997,Bryan1999,Osheaea2004}. To solve the
magnetohydrodynamical equations, we use a 2nd-order Runge-Kutta
temporal update of the conserved variables with the Local
Lax-Friedrichs (LLF) Riemann solver and a piecewise linear
reconstruction method. To ensure the solenoidal constraint on the
magnetic field, the divergence cleaning algorithm of
\citet{Dedner_ea_2002} is adopted
\citep{Wang_Abel_2008}. 

We implemented heating and cooling functions in the code that describe
both atomic and molecular heating and cooling processes (see 
\S\ref{sec:heatcool} for details). These functions take into account 
a density versus column density extinction relation similar to that of
\citet[henceforth, VLBT2013]{Van_Loo_Butler_Tan_2013}. 
As the temperature of the gas needs to
be calculated accurately, we use a {\textquotedblleft}dual energy
formalism{\textquotedblright} by solving the internal energy equation
as well as the total energy equation. The temperature is then
determined from the internal pressure when magnetic and kinetic energy
together exceed 0.999 the total energy, and from the total energy otherwise.

To track the evolution of properties of the clump, we use a scalar
value to differentiate between gas outside and inside the clump. We
set the scalar $S$ to 1 inside the clump and 0 outside. We added a
conservation equation in Enzo 2.0 to advect the scalar, i.e.,
\begin{equation} \label{eq:advection}
\frac{\partial(\rho S)}{\partial t} + \nabla .(\rho S \bf{v}) = 0,
\end{equation}
with $\rho$ the cell density and $\bf{v}$ the velocity.

We model a numerical domain of 256$^2\:$pc$^2$ which is covered by a
uniform grid of 1024$^2$, giving a grid cell size of 0.25~pc.  For the
fiducial model, two additional AMR grid levels are included, thus
increasing the effective resolution to 4096$^2$ with a grid cell size
of 0.0625~pc on the finest level.  This resolution is sufficient to
study the transition from subcritical to supercritical clouds and
clumps.

We use several criteria to determine refinement: a cell is refined when
there is a strong local gradient of variables (i.e., when the relative
slope $|q(i+1)-q(i-1))/q(i)|$ across variable $q$ at index $i$ exceeds 
0.4), when it is part of a shock front (defined by a relative pressure 
jump of $>0.33$), and when the local Jeans length is not covered by at 
least 4 cells \citep[needed to avoid artificial fragmentation][]{Truelove_ea_1997}.

\section{Heating and Cooling Functions}
\label{sec:heatcool}

We model the thermal properties of the ISM using a Photo-Dissociation
Region (PDR)-based method that follows and expands upon VLBT2013 and is 
detailed in the following subsections.

\subsection{Implementation of Thermal Processes}

Implementation into the {\it Enzo} code involves calculating the net heating
rate for a given cell:
\begin{equation} \label{eq:heating}
H = n_{\rm H}[\Gamma - n_{\rm H}\Lambda] \; \rm erg\, cm^{-3}\, s^{-1},
\end{equation}
where $\Gamma$ is the heating rate and $\Lambda$ is the cooling rate.
The net cooling rate introduces a cooling timescale: $t_{\rm cool}
\equiv E_{\rm int}/|H|$.  The internal energy of the gas is defined by
$E_{\rm int} = p/\left(\gamma - 1\right)$.

We adopt a mean particle mass of $\mu=2.33\:m_{\rm H}$ (valid for
molecular gas with 1 He per 10 H and ignoring contributions from other
species). For simplicity, this value of $\mu$ is adopted through the
simulation domain, i.e., even in the ambient, ``atomic'' medium.
 
The dynamics of the simulation (and the objects of main interest) are
dominated by gas at densities of $n_{\rm H} \gtrsim 10^2\:{\rm
  cm^{-3}}$, which correspond to equilibrium temperatures of $\sim
10$~K (details described below).  Thus, we adopt the value of
$\gamma=5/3$ for the entire simulation domain.  While this does not
account for the excitation of rotational and vibrational modes of
${\rm H_{2}}$ that would occur in shocks, we consider that this is the
most appropriate single-valued choice of $\gamma$ for our simulation
set-up, given our focus on the dynamics of the dense molecular gas.

The chosen values of $\gamma$ and $\mu$ set sound speeds of
$c_{s}=(\gamma k T / \mu m_{p})^{1/2} \approx 0.24(T/10~{\rm
  K})^{1/2}{\rm~km~s^{-1}}$.  
Since $t_{\rm cool}$ is often shorter than the hydrodynamical time,
the temperature and internal energy are sub-cycled and updated,
assuming constant density, until the hydrodynamical time step is
reached.  This is more computationally efficient as a method for
preventing excessive heating or cooling, than evolving all variables
on timesteps equal to the cooling or heating times.

\subsection{Density-Extinction Relation}

Simulating a $\sim{\rm kpc^{3}}$ region of a galactic disk, VLBT2013
found a monotonically increasing relation between density and average
(six orthogonal ray) column extinction, which defined an effective
visual extinction \citep{Glover_Mac_Low_2007}. This relation was
resolution-limited at high densities due to the effective visual
extinction being dominated by absorption within a single 0.5~pc cell.
For the simulations performed in this paper, we use a modified
extinction curve normalized to estimated values of the Warm Neutral
Medium (WNM): $A_{V} \simeq 0.01$~mag for $n_{\rm H} = 0.03 {\rm
  \,cm^{-3}}$ \citep{Wolfire_ea_2003}, GMCs: $A_{V} \simeq 1$~mag for
$n_{\rm H} = 100 {\rm \:cm^{-3}}$, and starless cores: $A_{V} \simeq
30$~mag for $n_{\rm H} = 10^{6} {\rm \,cm^{-3}}$.

\begin{figure}[H]
\begin{center}
\includegraphics[width=1.0\columnwidth]{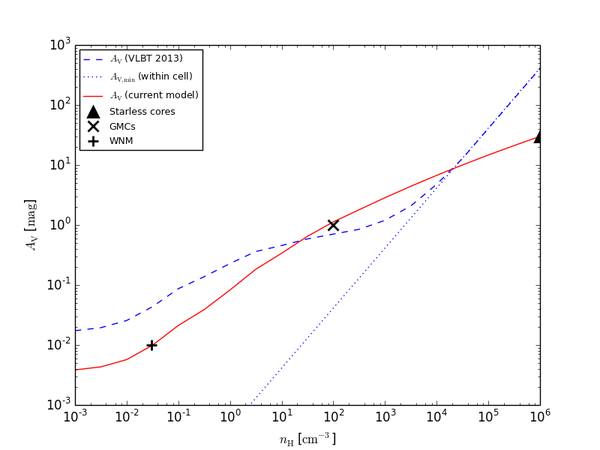}
\caption{\label{fig:density-extinction}
Average visual extinction as a function of density. The red solid line
represents the adopted $A_{V}$ versus $n_{\rm H}$ relation, based on
three observational constraints (see text). For comparison, the blue
dashed line represents the relation used by
VLBT2013, with the dotted line showing the
resolution limit due to extinction within the cell itself.
}
\end{center}
\end{figure}

To fit these constraints as well as retain the physical relationships
represented in the VLBT2013 curve, we ignore the effects of individual
cell extinction at high densities and instead perform a logarithmic
extrapolation from $n_{\rm H}=10^{3} {\rm \,cm^{-3}}$. This
intermediate curve is fitted to the normalization points via a smooth
scaling function, producing the final density-extinction relation (see
Fig.~\ref{fig:density-extinction}).

\subsection{Nonequilibrium PDR Heating and Cooling Rates}
\label{sec:PDRcompare}

We next utilize both \textsc{PyPDR}
\footnote{http://www.mpe.mpg.de/$\sim$simonbr/research\_pypdr/index.html} (described below)
and \textsc{Cloudy}
(version 13.02, last described in \citet{Ferland_ea_2013}),
photoionization simulation codes, to generate tables of
non-equilibrium heating and cooling rates as functions of density,
temperature and radiation field intensity. Our default value of FUV
radiation field intensity is $G_0=4$, following conditions developed
for the $\sim 4$~kpc molecular ring region of the inner Milky Way
(VLBT13). Then, given the $A_V$ vs. $n_{\rm H}$ relation described in
the last subsection, each value of density has a unique value of
received FUV intensity, allowing a 2D ($n_{\rm H}, T$) grid of heating
and cooling rates to be sufficient. However, to calculate this 2D grid
self-consistently does require calculation of arbitrary PDR models
with input density, temperature and radiation field, in order to
calculate species abundances correctly, especially of molecules like
${\rm H_{2}}$ and CO that have abundances set by FUV photons whose
propagation is affected by self-shielding.

To carry out these PDR models we primarily utilize \textsc{PyPDR}, which
is a minimal Python-based PDR code that self-consistently calculates
chemical, thermal, and molecular properties within a slab of gas
irradiated with FUV photons.  

\textsc{PyPDR} implements the same chemical network as in 
\citet{Rollig_ea_2007}, which includes reactions for H$_2$ formation, 
cosmic ray induced reactions, photodissociation (including 
self/mutual-shielding of H$_2$, CO and C) and gas-phase reactions.

The following heating and cooling rates are implemented:
H$_2$ pumping and line cooling \citep{Rollig_ea_2006}, 
H$_2$ formation \citep{Sternberg_Dalgarno_1989}, 
H$_2$ dissociation \citep{Jonkheid_ea_2004}, 
gas-grain heating/cooling \citep{Tielens_2005},
photoelectric heating and recombination cooling \citep{Bakes_Tielens_1994},
Ly-$\alpha$ cooling \citep{Sternberg_Dalgarno_1989},
optical Oxygen-6300~{\AA} cooling \citep{Sternberg_Dalgarno_1989},
heating by C-ionization \citep{Black_vanDishoeck_1987,Jonkheid_ea_2004},
cosmic ray heating \citep{Jonkheid_ea_2004},
line cooling by OI, CII, CI (fine structure), CO and $^{13}$CO(rotational)
calculated from the non-LTE excitation of OI, CII, CI, CO, and $^{13}$CO 
using an escape probability approach.
Data from the LAMDA database \citep{Schoeier_ea_2005} is used. 

While the \textsc{PyPDR} chemical network includes only $\sim$30 atoms
and molecules, it still performs well in benchmark tests, producing
results similar to larger PDR codes \citep{Rollig_ea_2007}.  However,
as \textsc{PyPDR} was developed for temperatures only up to
$\sim10^4$~K, we do not use it for higher temperature conditions.

\textsc{Cloudy}, on the other hand, follows a much larger number of 
species than \textsc{PyPDR} and can treat $T > 10^4$~K gas. Thus, we 
utilize it in this regime. However, for our purposes of defining 
non-equilibrium heating and cooling functions that utilize a two step 
process where high spectral resolution line self-shielding output is 
needed as a general input for the next PDR calculation, the public 
version of Cloudy does not automatically provide such output information. 
Thus we have adapted the PyPDR code of Bruderer for this purpose.

We set up the density-temperature parameter space for the arrays as
follows.  The density range is $n_{\rm H}=10^{-3}$ to $10^{6}\:{\rm
  cm^{-3}}$, in steps of 0.1 dex (91 values) while the temperature
spans from 2.7~K to $10^{5}$~K in steps of 0.046 dex (100
values). \textsc{PyPDR} was used to calculate the bulk of the rates,
from T = 2.7 to $10^{4}$~K, while \textsc{Cloudy} was used for T =
$10^{4}$ to $10^{5}$~K.

The procedure for both PDR codes generally follows that of
VLBT2013, which used \textsc{Cloudy} version
8.02 and was based off of \citet{Smith_ea_2008}.  Any code-specific
differences will be mentioned in the relevant sections.

First, the unextinguished local interstellar radiation field (ISRF)
with $G_{0}=4$ is incident on an absorbing slab of gas with
abundances, metallicities, and dust resembling that of the local ISM.
We include the cosmic microwave background radiation as well as a
background of cosmic rays with primary ionization rate of $\zeta = 1.0
\times 10^{-16}\:{\rm s^{-1}}$.  The column density of the gas slab is
determined by the previously described density-extinction relation and
the linear relation between column density and visual extinction
($A_{V}= 5.35\times 10^{-22} N_{\rm H}$~mag).  For a given density, a
PDR model is calculated through a slab with column corresponding to
the particular density.  In our case, the density of the slab also
follows this given density, as the extinction should be dominated by
the local density for GMC regions.  Note, in VLBT13, the density of
the slab was fixed at $n_{\rm H} = 1 {\rm \,cm^{-3}}$.

At the depth of the specified column (i.e., the final cell of the PDR
model) the temperature of a parcel of gas is then varied, with heating
and cooling rates calculated for the specific temperature. The
calculations are repeated for the entire temperature range, yielding
the temperature and density dependent heating and cooling functions.

\begin{figure}[h!]
\begin{center}
\includegraphics[width=1.0\columnwidth]{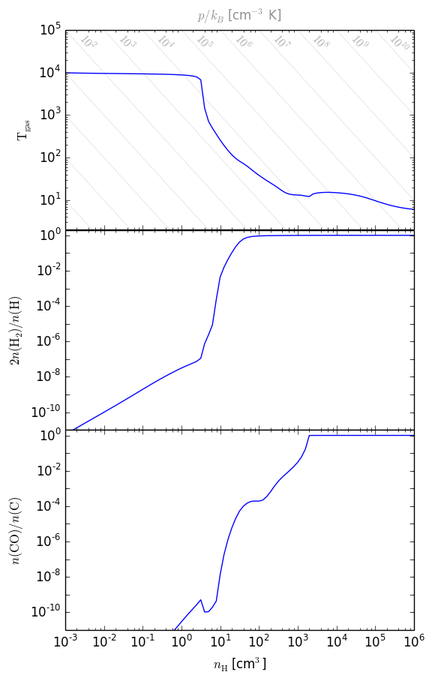}
\caption{\label{fig:Teq-models}
({\it top}) 
\textsc{PyPDR} equilibrium temperature as a function of density.  The density
corresponds to a value of $A_{\rm V}$ from from Fig.~\ref{fig:density-extinction}.
Details are discussed in \S\ref{sec:PDRcompare}.
Lines of constant pressure are plotted in gray to more easily show regions of thermal
instability.  
({\it middle}) ${\rm H_{2}}$ fraction as a function of
density.  
Hydrogen becomes essentially fully molecular at densities above
$n_{\rm H}\simeq 80\:{\rm cm^{-3}}$.  ({\it bottom}) CO fraction as a
function of density.  The carbon becomes fully molecular in the form
of CO at densities above $n_{\rm H}\simeq2\times10^{3}{\rm
  \,cm^{-3}}$.  }
\end{center}
\end{figure}

In PyPDR, the code 
allows self-consistent calculation of general, nonequilibrium heating
and cooling rates given species abundances set by equilbrium PDR
conditions for given extinction, density and equilibrium temperature.
The key PyPDR results for $T_{\rm eq}$, ${\rm H_{2}}$, and ${\rm CO}$
abundances are shown in Fig.~\ref{fig:Teq-models}.

Arrays of heating and cooling rates were created using both \textsc{PyPDR}
($T=2.7$~K to $10^{4}$~K) and \textsc{Cloudy} ($10^{4}$~K to
$10^{5}$~K), then smoothly joined along the temperature dimension
using the function:
\begin{equation} \label{eq:fermidirac2}
R(T)=10.0^{\log_{10} \left[ R_{C}(T) \alpha (T) \right] + \log_{10} \left[ R_{P}(T) (1.0-\alpha (T)) \right]}
\end{equation}
where $R(T)$ is the final, smoothly combined rate, calculated from $T$
the gas temperature, $R_{C}$ the \textsc{Cloudy} rate, $R_{P}$ the
\textsc{PyPDR} rate, and the Fermi-Dirac smoothing function:
\begin{equation} \label{eq:fermidirac1}
\alpha(T) = \frac{1} {1+\exp{\left[ -10.0 (\log_{10} T - 4.0) \right] }}.
\end{equation}
This joined the functions at $T=10^{4}$~K, where
good agreement still occurred between the models.

\begin{figure}[h]
\begin{center}
\includegraphics[width=1.0\columnwidth]{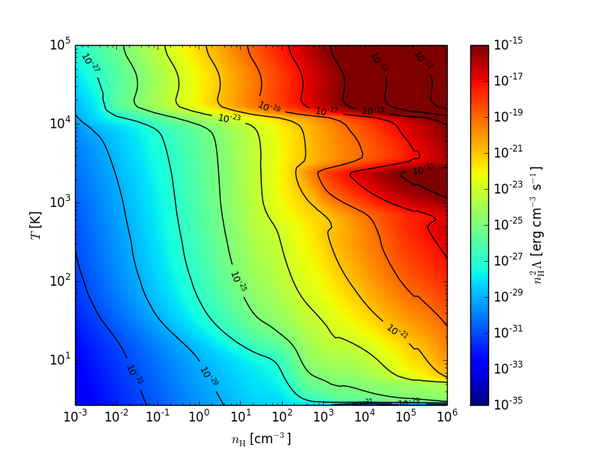}
\includegraphics[width=1.0\columnwidth]{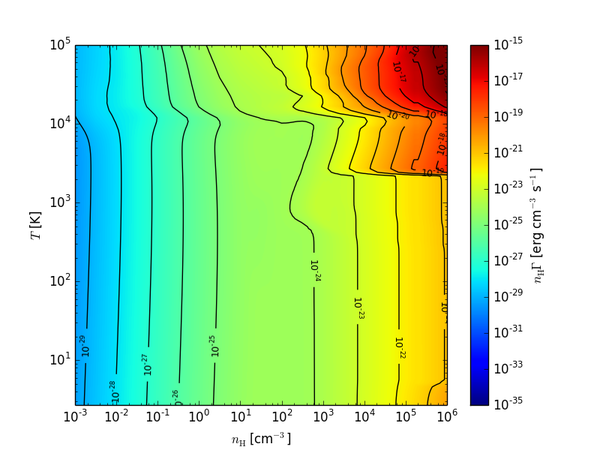}
\includegraphics[width=1.0\columnwidth]{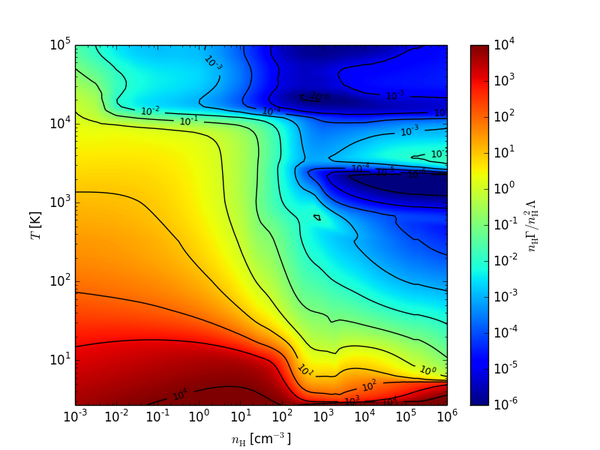}
\caption{\label{fig:extrap-maps}
Density and temperature dependent interpolated arrays of ({\it top})
cooling, ({\it middle}) heating, and ({\it bottom}) ratio of
heating/cooling. The contours show constant rates ({\it top} and {\it
  middle} panels) and ratios ({\it bottom} panel), e.g., in the ratio
map, the $10^{0}$ contour represents the equilibrium temperature.}
\end{center}
\end{figure}

From the resulting final arrays of heating rates and cooling rates, a
bilinear interpolation is performed to derive rates for any density
and temperature combination. The final, combined 2D interpolation
plots for cooling, heating, and net heating as functions of density
and temperature are displayed in Fig.~\ref{fig:extrap-maps}.  These
plots show the total cooling, heating, and net rate at all densities
and temperatures.

Note, that from $T = 10^5$ K up to $T = 10^8$ K, we ignore heating and 
adopt the cooling rates derived
by \citet{Sarazin_White_1987}.  Values beyond the array domain adopt
the limiting values.  Thus, for any cell in our Enzo simulation, the
density and temperature are read in and cooling and heating rates are
returned.

\subsection{Heating and Cooling Components}

\begin{figure}[h]
\begin{center}
\includegraphics[width=1.0\columnwidth]{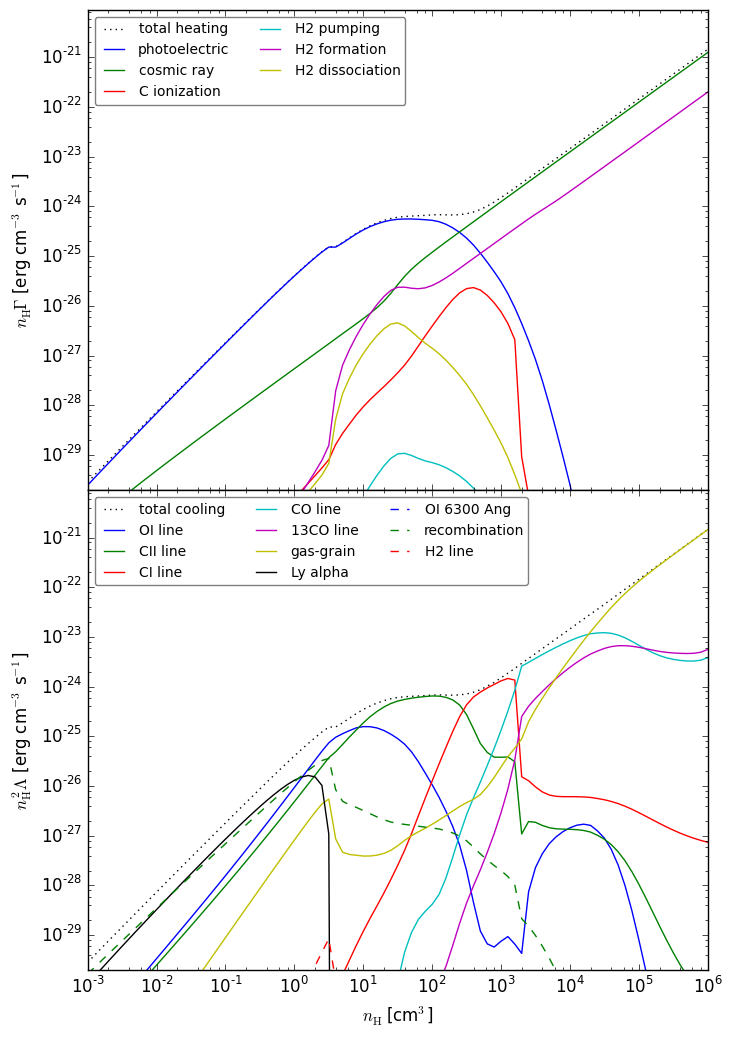}
\caption{\label{fig:component-breakdown}
Component breakdown of the main cooling ({\it top}) and heating ({\it
  bottom}) rates per unit volume as a function of density at the
respective equilibrium temperatures (given in
Fig.~\ref{fig:Teq-models}).}
\end{center}
\end{figure}

A breakdown of specific heating and cooling components at the
equilibrium temperature is shown in
Fig.~\ref{fig:component-breakdown}.  Photoelectric heating of dust
grains is the dominant heating source for low-density gas up until
$n_{\rm H} \sim 10^{2}\:{\rm cm^{-3}}$.  Above this density, the
higher dust extinction blocks external FUV photons and thus reduces
photoelectric heating.  The ubiquitous flux of cosmic rays then
becomes the main heating component in high-density gas.  ${\rm H_{2}}$
formation also contributes as the cloud becomes fully molecular.

The main coolants in the low density, ionized/atomic region ($n_{\rm
  H} < 1\:{\rm cm^{-3}}$) are Ly-$\alpha$ and Hydrogen recombination
lines.  As density increases, various atomic lines (OI, CII, CI)
become dominant coolants.  These species inelastically collide with H
and He, exciting internal degrees of freedom and subsequently decaying
through photon emission.  Molecular lines (CO, $^{13}$CO) provide
large contributions in cooling as density increases, temperature
decreases, and the gas reaches high levels of molecular abundance.  At
the highest densities ($>10^{4}\:{\rm cm^{-3}}$), gas-grain cooling
dominates as collisions between dust grains and gas molecules lead to
emission of infrared photons from the decay of lattice vibrations.

\vspace{10mm}

\subsection{Observational Diagnostics}
\label{sec:heatcool-diagnostics}

In addition to providing a better understanding of the dominant
physical processes occurring at different densities and temperatures,
the heating/cooling component breakdown also enables the creation of
observational diagnostics in the form of line emissivities.  Here we
focus on a preliminary investigation into high-$J$ CO to see if
they are good diagnostics of shocks arising from
cloud collisions.  The PDR-derived cooling data include contributions
from the first 40 rotational lines for both $^{12}$CO and
$^{13}$CO. Similar to the method of creating the cooling and heating
functions, tables of density and temperature dependent emissivities
were compiled to allow calculation of observable quantities in the
form of integrated intensity maps and spectra.

Via post-processing, integrated intensity maps can be derived from the
volume emissivity functions coupled with simulation outputs. 
We assume a fixed 1~pc thickness of the simulation volume for
calcuation of these maps. Given this simplistic, highly-idealized 2D
geometry of cloud structures presented in this initial paper, for
simplicity we do not calculate radiative transfer of the emissivities
from each cell, but simply sum their contributions as if their
emission reached us with negligible attenuation. However, note that
the emissivities of lines from \textsc{PyPDR} do already account for
an idealized cloud optical depth via an escape probability formalism
through the PDR layer (see \S\ref{sec:PDRcompare}).
Further detailed study of observational diagnostics of cloud-cloud
collisions based on 3D simulations and including full radiative
transfer will be deferred to a future paper.

For ease of comparison with Galactic clouds, we assume a fiducial 
cloud distance of $d=3\:\rm{kpc}$ and depth of $z = 1\:{\rm pc}$. We 
use the line volume emissivities derived from the PyPDR to determine 
an integrated intensity for each cell in the simulation. 
Integrated intensities are derived via:
\begin{equation} 
    I = \int I_{\nu}{\rm d}\nu = \frac{2k}{\lambda^{2}}\int T_{\rm mb} {\rm d}\nu.
\end{equation}
where $I_{\nu}$ is the specific intensity, $\lambda$ is the wavelength 
of the chosen molecular line, and $T_{\rm mb}$ is the main beam 
temperature.
Changing variables from $\nu$ to $v$ and substituting, we have
\begin{equation} 
    \int T_{\rm mb} {\rm d}v = \frac{\lambda^{3}}{2k} I = \frac{\lambda^{3} j V}{8 \pi k d^{2} \Omega}.
\end{equation}
where $j$ is the volume emissivity, $V$ is the cell volume, and 
$\Omega$ is the solid angle subtended by the cell. 

While values of $z$ and $d$ are assumed in order to provide 
some observational outputs, the intensity maps can be scaled for 
any desired thickness, given the optically-thin assumption. Integrated 
intensity maps of CO lines and line ratios with rotational excitations 
$J$=2-1, 3-2, and 8-7 using this method are presented and discussed in 
\S\ref{sec:diagnostics}.

In addition to integrated intensity maps, spectra of the corresponding
observational volumes can be created, simply by plotting the
distribution of specific intensity as a function of line of sight velocity.
Synthetic spectra of the $^{13}$CO($J$=2-1), $^{13}$CO($J$=3-2), and 
$^{12}$CO($J$=8-7) lines for an isolated GMC and a GMC collision case, 
viewed along sight lines within the 2D simulation
plane, are compared. Velocity gradients derived from these spectra
are described as well in \S\ref{sec:diagnostics}.

\section{Results}
\label{sec:results}

\subsection{Out-of-plane magnetic fields}

Here we assume the magnetic fields are orientated orthogonal to the
2D simulation plane and thus the collision velocity. Using the virial
theorem, \citet{Chandrasekhar_Fermi_1953} showed that magnetic fields
support the cloud preventing gravitational collapse if the average
magnetic field strength exceeds
\begin{equation} \label{eqn:Bcrit}
    B_{\rm crit} = 2 \pi R \bar{\rho} G^{1/2},
\end{equation}
where $\bar{\rho}$ is the average density of the cloud. Note this assumes
the external magnetic field to be negligible.  We systematically vary
the magnetic field strength to probe the sub- and supercritical
regimes and to understand the transition from sub- to supercritical.

\subsubsection{Isolated cloud}\label{magnetic_criticality}

The simplest case to consider is an isolated cloud. For our adopted
parameters (see GMC 1 values of Table~\ref{tab:GMC-properties} and
runs 1.A.x in Tab.~\ref{tab:all_runs}), the critical magnetic field is
$27.9\:{\rm \mu G}$.  We initialize the cloud with a uniform
out-of-plane magnetic field, sampling values of 
10, 20, 27.9 and
40~${\rm \mu G}$ and setting the ambient field to 
10~${\rm \mu G}$. We also carry out an unmagnetized simulation.
Cloud evolution is followed for 10~Myr, more than 2 freefall
times, $t_{\rm ff} = (3 \pi / [32 G \rho])^{1/2}$, which is
$\simeq4.35$~Myr for the adopted cloud values.  Note that this
expression for the free-fall time is for a uniform sphere and not for
an infinite, uniform cylinder. This is an intentional choice to use a
common definition, as we will extend this work to 3D in future
studies.

\begin{figure}[t]
\begin{center}
\includegraphics[width=1.0\columnwidth]{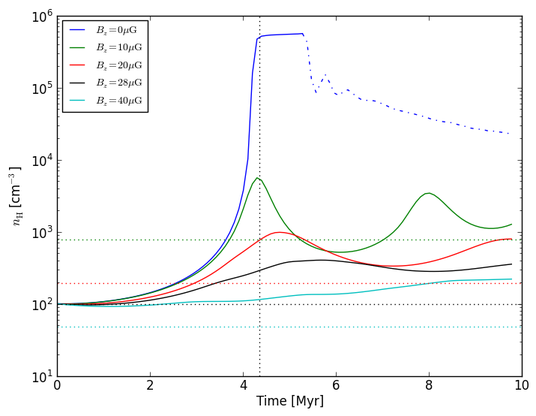}
\caption{\label{fig:crit-test-summary}
Average density of the cloud over time for initial $B$ strengths of 0,
10, 20, 28, and 40~$\rm \mu G$.  The blue dash-dotted line marks when
the $B=0~\rm \mu G$ case is affected by numerical effects. Critical field strength occurs
at $B_{\mathrm{crit}}=28\:{\rm \mu G}$.  The dotted vertical line
denotes the freefall time of the cloud.  The dotted horizontal lines
show $\rho_{\rm crit}$ predicted for each model.}
\end{center}
\end{figure}

Figure~\ref{fig:crit-test-summary} shows the evolution of the average
cloud density for each model, tracked using the advective scalar
method (see \S\ref{sec:num_method}).

For the unmagnetized, pure hydrodynamical model, the line-mass of the
cloud exceeds the critical value $2\sigma^2/G$ and thus collapses
unimpeded. All of the cloud mass ends up into a single cell in about a
free-fall time and then continues to slowly accrete mass from the
external medium (as seen in the shallow flat slope of the mean density
at 5~Myr).  Note that after $t \simeq 6 {\rm Myr}$, the cloud material
is no longer tracked well, likely the result of a numerical artefact
due to numerical diffusion. Note that we track the cloud (or clump,
as in later cases) by advecting a scalar field $S$ which we set to 1
inside the cloud and 0 outside. The cloud is defined by material with
$S>0.5$. By the time the cloud collapses, most of the mass is within a
small number of grid cells. As evolution continues, cloud and ambient
material mixes so that the scalar is now below the defined value and
its mass is no longer accounted for.

If magnetic fields are present, the magnetic pressure supports the
cloud against gravitational collapse. The oscillation is due to the
transition towards an equilibrium density and magnetic field
distribution. 

For magnetic field values below the critical value, the cloud is
initially supercritical and thus collapses. The average density
follows the same evolution as for the pure hydrodynamic
model. However, the clouds do not collapse completely, even for field
strengths close to the external magnetic field value. Actually, as
long as a magnetic field is present, the collapse of the cloud is
impeded. This can be easily understood from conservation of magnetic
flux and mass in a 2D geometry. Both the average magnetic field and
density are proportional to $1/R^2$. The critical magnetic field,
however, is proportional to $1/R$ as $B_{\rm crit} \propto R \rho$
(see Eq.~\ref{eqn:Bcrit}). Thus, as the cloud collapses, the average
magnetic field in the cloud increases faster than the critical
magnetic value. While the initial magnetic field is too weak to
support the cloud, subsequent contraction causes the field to
eventually become strong enough to halt collapse. Thus, the cloud
transitions from a supercritical regime to a subcritical one.

Using the initial line mass and magnetic flux of the cloud, the mean
density at which the transition from supercritical to subcritical
occurs, is given by
\begin{equation} \label{eqn:rho_crit}
\rho_{\rm crit} = \frac{4 \pi m_{l}^{3} G}{\Phi^{2}}.
\end{equation}
For $B=10\:{\rm \mu G}$ and $n_{\rm H} = 100\:{\rm cm^{-3}}$, we find
$n_{\rm crit} = 776\:{\rm cm^{-3}}$.
Figure~\ref{fig:crit-test-summary} indeed shows that the gravitational
collapse oscillates close to this value for $B = 10\:{\rm \mu G}$,
gradually settling towards an equilibrium state.  The final densities
are slightly higher than predicted, presumably due to external
pressure from the ambient, magnetized, infalling gas.

This result is specific to the 2D cylindrical geometry adopted
here. For a spherical cloud, the critical magnetic field strength is
given by \citet{Chandrasekhar_Fermi_1953}
\begin{equation} \label{eqn:2D_B_crit}
B_{\rm crit} \approx 2.5 \pi R \bar{\rho} G^{1/2}.
\end{equation}
Note the similarity with the expression of the critical value for a
cylindrical cloud (eq.~\ref{eqn:Bcrit}).  If the cloud is initially
supercritical, we can assume nearly isotropic collapse. Then, the
density is proportional to $1/R^3$, so that $B_{\rm crit} \propto
1/R^2$. Because of magnetic flux freezing in ideal MHD, the mean
magnetic field of the cloud is proporptional to $1/R^2$. This means
that the critical magnetic field and the mean magnetic field have the
same proportianlity and the cloud remains supercritical during the
collapse.

It is clear that, if high gravitational collapse is to be achieved in
2D, flow along field lines (along the direction of collapse) or flow
through field lines (due to, e.g., ambipolar diffusion or turbulent
reconnection) must occur. Alternate field geometries are discussed in
later sections.

\subsubsection{Isolated cloud with embedded clump}

\begin{figure}[t]
\begin{center}
\includegraphics[width=1.0\columnwidth]{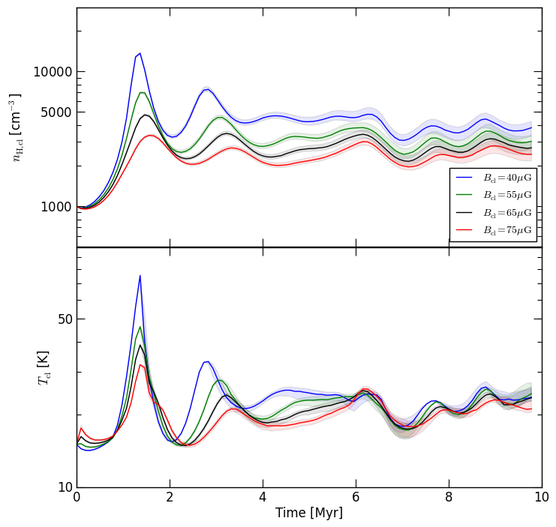}
\caption{\label{fig:vary-Bclp}
Average clump density ({\it top panel}) and temperature ({\it bottom
  panel}) versus time for constant GMC magnetic field $B_{1}= 40\:{\rm
  \mu G}$ and varying $B_{\rm cl}= 40, 55, 65, 75~{\rm \mu
  G}$. Critical field strength occurs at $B_{\rm cl}= 65~{\rm \mu
  G}$.  The solid line represents the average value in the clump
defined by the scalar value $S>0.5$, while the shaded regions show the
averages between $S>0.25$ and $S>0.75$.  This convention for clump
definition is followed throughout the remainder of the paper.  
}
\end{center}
\end{figure}

\begin{figure}[t]
\begin{center}
\includegraphics[width=1.0\columnwidth]{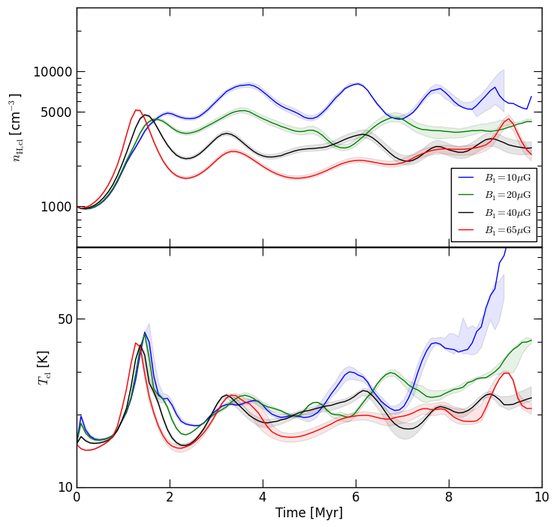}
\caption{\label{fig:vary-B1}
Average clump density ({\it top panel}) and temperature ({\it bottom
  panel}) over time for constant clump magnetic field $B_{\rm cl}=
65\:{\rm \mu G}$ and varying GMC field strength as $B_{1}= 10, 20, 40,
65~{\rm \mu G}$. Critical field strength occurs at $B_{1}= 40~{\rm \mu
  G}$.
}
\end{center}
\end{figure}

We now embed a clump within the cloud discussed in the previous
section.  The critical magnetic field for the clump is $\simeq
65\:{\rm \mu G}$, while the critical value for the cloud has increased
to $\simeq 40~{\rm \mu G}$ (as the average density of the cloud is
higher with the embedded clump).  We examine the effect of various
magnetic field strengths in the cloud and clump. First, we keep the
the clump magnetic field constant and vary the cloud value. This tells
us more about the evolution of an equilibrium clump in a sub- or
supercritical cloud. Then, we keep the cloud magnetic field constant
while varying the clump value. These correspond to runs 1.B.x in
Tab.~\ref{tab:all_runs}. Although we are restricted with our cloud and
magnetic field geometry, these results are useful for understanding
more complex simulations.  Results are shown in
Figures~\ref{fig:vary-Bclp} and ~\ref{fig:vary-B1}.

For a constant $B_{\rm GMC}$ near its critical value, the evolution of
the clump is entirely determined by the ratio of its gravitational and
magnetic energy (its thermal energy is negligible). Using
eq.~\ref{eqn:rho_crit}, we find that, for $B_{\rm cl} = 40\:{\rm \mu
  G}$, the average density of the clump increases by a factor of
2.7. However, Figure~\ref{fig:vary-Bclp} shows an increase twice this
value. This is due to the initial contraction of the cloud between 0-2
Myr as it tries to set up an equilibrium distribution. After this
initial adjustment phase, the average density of the clump drops to a
few times the initial value as expected. For higher initial magnetic
fields in the clump the density increases by a smaller factor, as also
expected from eq.~\ref{eqn:rho_crit}.

For a constant $B_{\rm cl}$, changes in the average clump density are
driven by external pressure from the surrounding GMC material. This
external pressure is relatively larger for the lower GMC magnetic
fields.  In these cases, the GMC is initially supercritical and starts
to contract gravitationally. The average density of the clump
increases maximally by a factor of $\sim 5$. The clump magnetic field
is strong enough to resist gravity but the clump is further compressed
to higher densities because the external pressure contributes
non-negligibly to the gravity. For stronger GMC magnetic fields,
however, the density of the clump is not increasing because of the
pressure exerted by the GMC. Instead, the clump is initially no longer
subcritical. The external (i.e., GMC) magnetic field is not negligible
and should be taken into account when deriving the critical value.
For high GMC magnetic fields, the critical value of the clump is
actually greater than $65~{\rm \mu G}$, and thus it initially
collapses gravitationally.  However, it is not highly supercritical so
the density increase is quite modest. At the same time, GMCs with
higher magnetic fields expand after 2-2.5~Myr (see previous
section). The external magnetic field then decreases, as well as the
critical magnetic field of the clump. This results in re-expansion of
the clump to near its initial value. Our results suggest that
increasing external pressures is a possible way to trigger a
sub-to-supercritical transition.

\subsubsection{Colliding clouds: head-on collisions}

A significant source of additional pressure can be provided by ram
pressure of cloud collisions and the resulting thermal and magnetic
pressure released in shocks. The ram pressure depends on the relative
collision speed, ${v_{\rm rel}}^2$. We investigate different collision
speeds, i.e., $v_{\rm rel} = 5, 10, 20, {\rm and}~40~{\rm km\ s^{-1}}$
(see runs 1.C.1.x in Table~\ref{tab:all_runs}).

\begin{figure*}[t]
\begin{center}
\includegraphics[width=2.0\columnwidth]{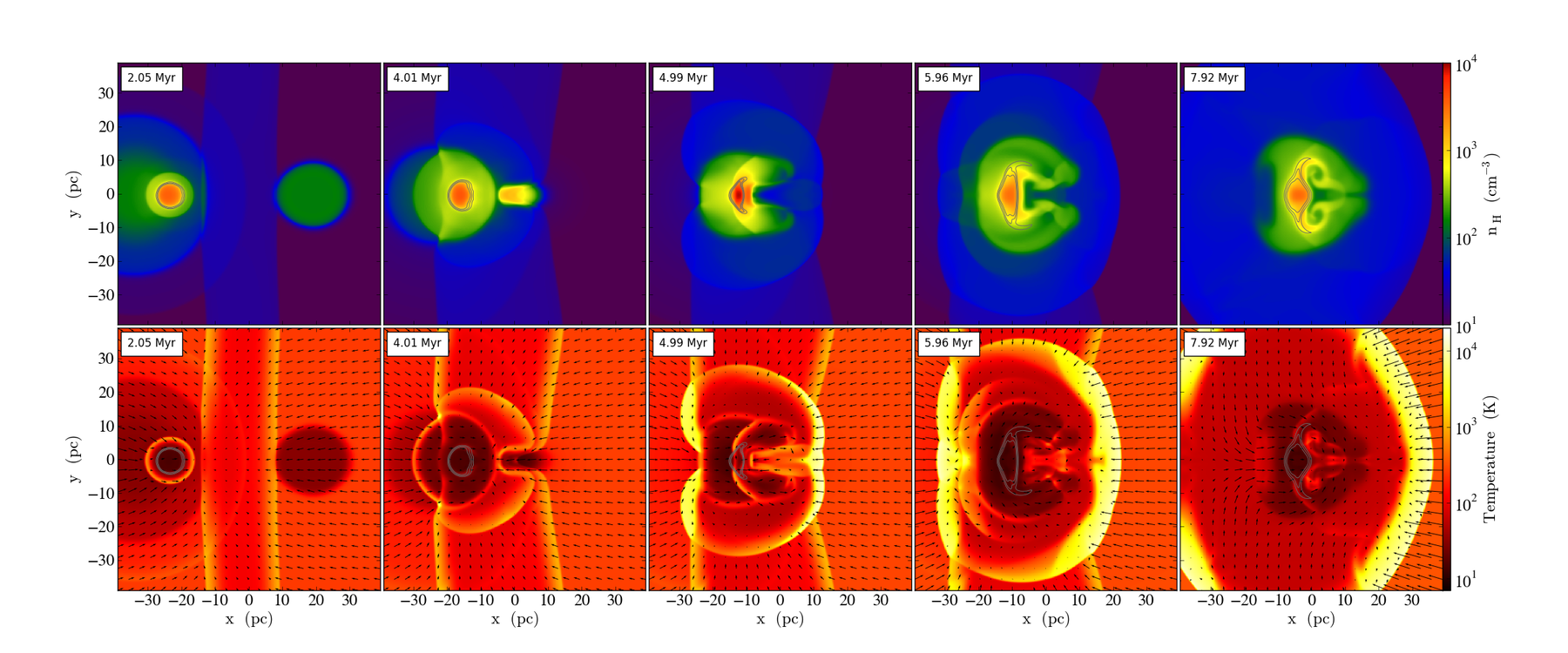}
\caption{\label{fig:colliding}
Evolution of clouds colliding head-on (zero impact parameter) with
snapshots shown at 2.05, 4.01, 4.99, 5.96, and 7.92~Myr (see run 1.C.0
in Table~\ref{tab:all_runs}.)  Here, magnetic fields are near critical
values and directed out-of-plane.  $B_{\rm cl}=65\:{\rm \mu G}$,
$B_{1}=40\:{\rm \mu G}$, and $B_{0}=10\:{\rm \mu G}$.  ({\it top row})
Maps of $n_{\rm H}$ and ({\it bottom row}) temperature, with black
vectors representing velocity are shown.  The advective scalar
defining the clump is shown by grey contour lines, representing the
scalar value $S=0.25$, 0.5, and 0.75.  }
\end{center}
\end{figure*}

\begin{figure}[t]
\begin{center}
\includegraphics[width=1.0\columnwidth]{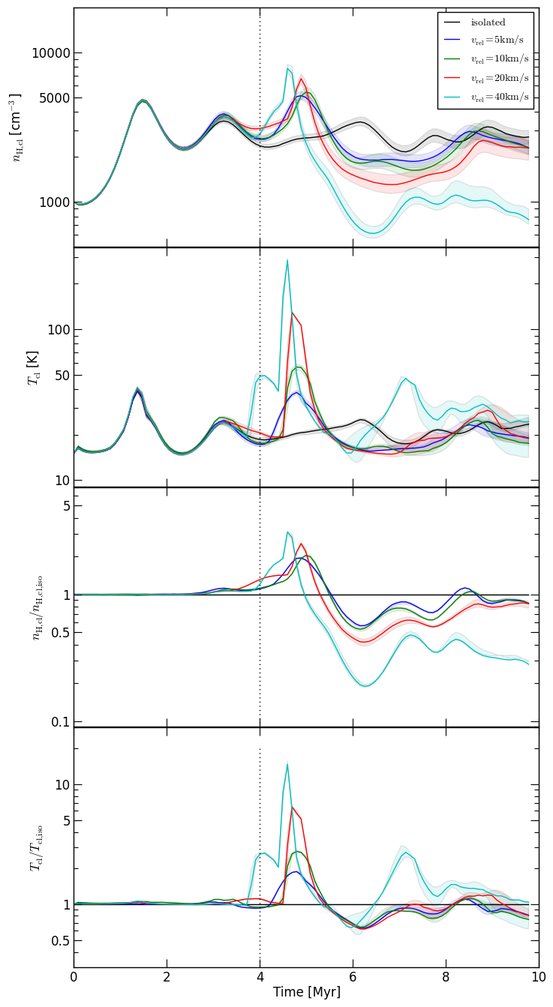}
\caption{\label{fig:summary-Bz_col}
Average clump density ({\it Top panel}) and temperature ({\it 2nd
 panel}) over time comparing the effects of collision velocity for
out-of-plane field geometries (see runs 1.C.1.x).  Here, out-of-plane 
magnetic fields are near critical
($B_{\rm cl}=65\:{\rm \mu G}$, $B_{1}=40\:{\rm \mu G}$, and $B_{0}=10\:{\rm \mu G}$).
Collision velocities $v_{\rm rel}= 5,
10, 20, 40 {\rm km\ s^{-1}}$ are shown, along with the evolution of
the isolated GMC with clump.  The ratios of the colliding cases compared 
to the isolated case for average density ({\it 3rd panel}) and
temperature ({\it bottom panel}) are also shown. }
\end{center}
\end{figure}

Density and temperature slices at different stages of the evolution
are shown for $v_{\rm rel}=10\:{\rm km\: s^{-1}}$ in
Figure~\ref{fig:colliding}. 
The two clouds are initially separated such that
the collision occurs at 4~Myr, which allows for an initial redistribution
of the density in the cloud (see Figures~\ref{fig:summary-Bz_col} and
\ref{fig:summary-Bz_col_b}).  The cloud-cloud collision compresses the
clouds and the clump, leading to higher densities. The collision also
gives rise to many shocks propagating through the clouds. Such shocks
contribute to raising the pressure 
around the clump. High-temperature shock fronts are present within the
otherwise cold ($\sim 15$~K) clouds. The magnitude of the magnetic field also
increases as material is compressed.  This increase in magnetic
pressure prevents the clump from collapsing completely, even with the
additional external pressure of the collision.

Figure~\ref{fig:summary-Bz_col} shows the evolution of the average
clump density and temperature for different collision speeds compared
to the non-colliding case.  Note that, for 
$v_{\rm rel} = 20$ and $40~{\rm km\ s^{-1}}$, the collisions were
performed in the reference frame of the clump, to avoid high flow
velocity induced numerical diffusion effects that can have modest
effects on clump boundary definitions, mostly affecting measurement of
clump temperature.

Due to the utilized set-up, a collision front between the low-density
ambient envelopes arises in between the clouds. Before the clouds
directly interact, they are being influenced somewhat by this high
pressure post-shock collision region. However, the pressure here is
much less than the ram pressure resulting from the GMC-GMC collision,
given the factor of 10 difference in GMC to ambient density. The
effects of the shocked ambient medium on the clump can be seen in
Fig.~\ref{fig:summary-Bz_col}, at $t\lessapprox4$~Myr. Both density
and temperature are affected, more noticeably at 20 and $40~{\rm
  km\ s^{-1}}$, but the ensuing GMC collision dominates the subsequent
clump evolution. These effects due to the shocked ambient medium can
be seen in the density and temperature evolution for colliding cases
in subsequent runs, discussed below.

As expected, the density and temperature of the clump resulting from
the GMC collision increase with collision speed, as higher velocities
induce stronger shocks with larger compressions. However, even for
$v_{\rm rel} = 40~{\rm km\ s^{-1}}$, the increases in clump density
are only modest: about a factor of 2 to 3 times greater than the
isolated case. 
Of course, some of this is due to the specific geometry we adopt
here. For other cloud geometries, e.g., spherical clouds in 3D, and
magnetic field geometries, e.g., more parallel to collision
velocities, this extra pressure may yet be sufficient to trigger the
transition from sub- to supercritical. The collision models do show
larger excursions in clump mean temperatures, which would be expected
to have an impact on astrochemical processes in the clump.

\subsubsection{Colliding clouds: off-axis collisions}
\label{sec:bparam}

\begin{figure*}[t]
\begin{center}
\includegraphics[width=2.0\columnwidth]{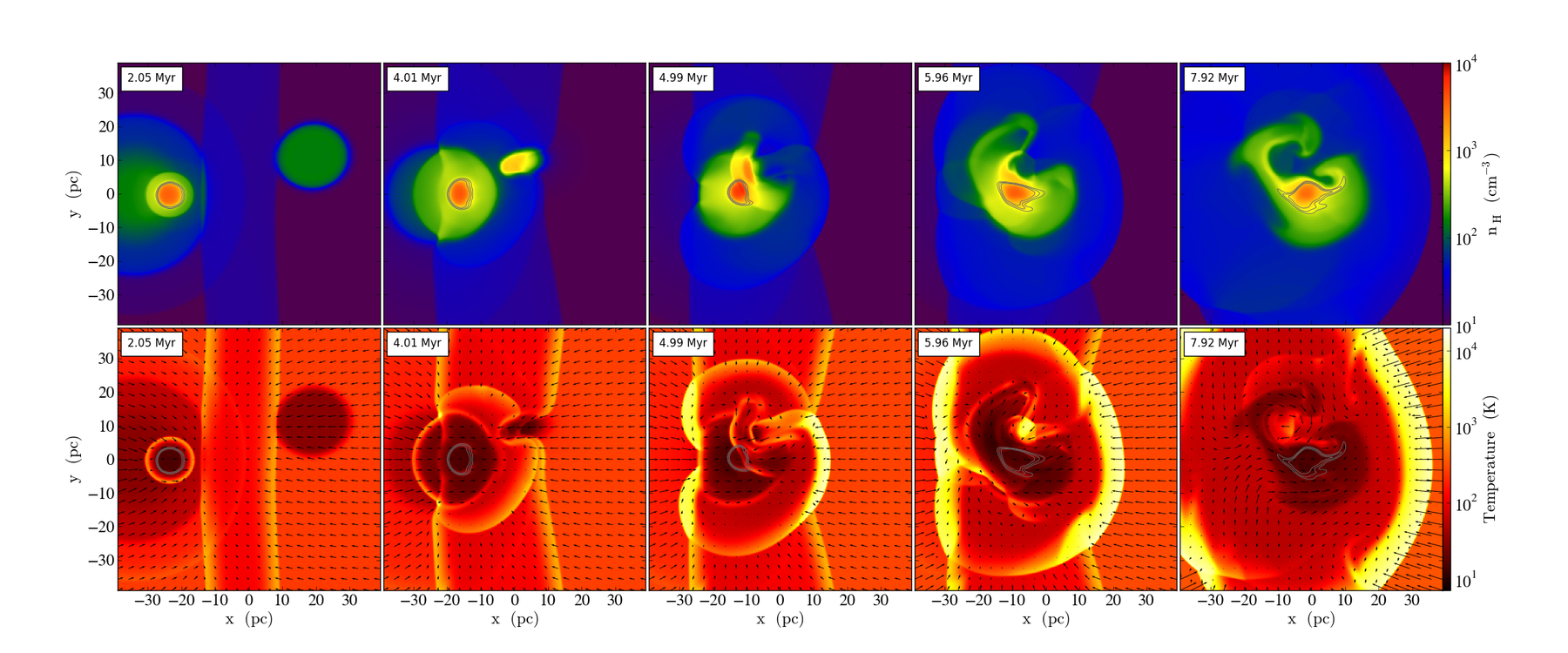}
\caption{\label{fig:map-Bz_b}
Time evolution of colliding clouds at 2.05, 4.01, 4.99, 5.96, and 7.92~Myr
with $b = 0.5 R_{1}$ (see run
1.C.2.2 in Table~\ref{tab:all_runs}.) 
Here, out-of-plane magnetic fields are near critical
($B_{\rm cl}=65\:{\rm \mu G}$, $B_{1}=40\:{\rm \mu G}$, and $B_{0}=10\:{\rm \mu G}$).
({\it top row}) Maps of $n_{\rm H}$ and ({\it bottom}) maps of temperature
with black vectors representing velocity are shown.  The advective scalar
defining the clump is shown by grey contour lines, representing the
scalar value $S=0.25$, 0.5, and 0.75.  }
\end{center}
\end{figure*}

\begin{figure}[t]
\begin{center}
\includegraphics[width=1.0\columnwidth]{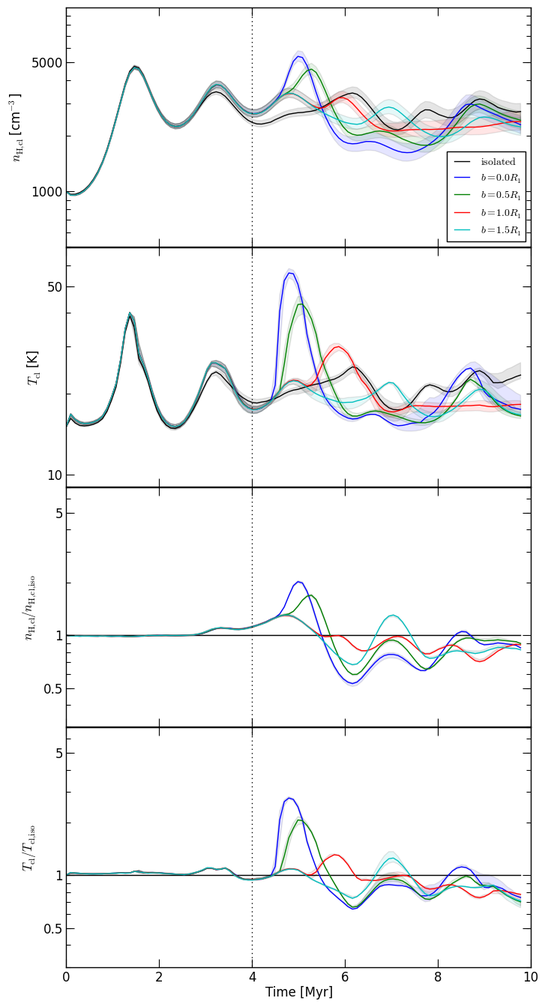}
\caption{\label{fig:summary-Bz_col_b}
Average clump density ({\it Top panel}) and temperature ({\it 2nd
 panel}) over time comparing the effects of impact parameter for
out-of-plane field geometries (see runs 1.C.2.x).  Here, out-of-plane 
magnetic fields are near critical
($B_{\rm cl}=65\:{\rm \mu G}$, $B_{1}=40\:{\rm \mu G}$, and $B_{0}=10\:{\rm \mu G}$).
Impact parameters of $b = 0.5, 1.0$, and $1.5 R_{1}$ were explored.  
The ratios of the colliding cases compared to the isolated case for average 
density ({\it 3rd  panel} and temperature ({\it bottom panel}) are also shown. }
\end{center}
\end{figure}

Off-axis collisions, in which the impact parameter was varied, were
also explored.  GMC 2 was placed at different perpendicular distances,
$b$, from GMC 1's line of symmetry and the mean density of the clump
material was tracked over 10 Myr. Figure~\ref{fig:map-Bz_b} displays
the morphology of the collision for $b=0.5 R_{1}$.  The clouds
interact at $\sim 4$ Myr in an asymmetric manner, creating filamentary
structures and imparting angular momentum on the coalesced structure.
Compared with the on-axis head-on collisions, the resulting structures
are morphologically more filamentary but the level of gravitational
contraction is roughly equivalent.  The lack of complete gravitational
contraction is expected because of the flux-freezing limitation of
out-of-plane fields described above.  In addition, any angular
momentum in the final structure also helps to support the clump,
further reducing the degree of contraction.

The average clump densities for various impact parameters are compared
in Figure~\ref{fig:summary-Bz_col_b}.  Collisions at $\sim 4$ Myr show
varying factors of density increase, with higher average densities for
smaller values of $b$ (more direct collisions).  As with the case of
head on collisions, clump densities are only increased by at most a
factor of a few.

\subsection{In-plane magnetic fields}

The primary inhibitor of complete collapse in the out-of-plane
($B_z$) magnetic field runs is flux freezing, i.e., gas cannot move
across magnetic field lines. Therefore, in this section, we change the
direction of the magnetic field from orthogonal to the plane to be
within the plane. Contrary to the out-of-plane field models where the
magnetic field value is higher inside the cloud than outside it, we
assume a uniform magnetic field across the cloud and external medium.
For such clouds, gravitational collapse proceeds preferentially along
the magnetic field lines.  While forces supporting the cloud are much
greater perpendicular to the magnetic field lines,
\citet{Tomisaka_2014} showed that a uniform in-plane field geometry
can yield magnetically subcritical configurations for infinite
cylinders.  The maximum line mass was evaluated as
\begin{eqnarray} \label{eq:tomisaka}
\lambda_{\rm max} \simeq 22.4 \left( \frac{R_{0}}{0.5\:{\rm pc}} \right)  \left( \frac{B_{0}}{10\:{\rm \mu G}} \right) M_{\odot}\:{\rm pc}^{-1} 
\nonumber \\ + 13.9 \left( \frac{c_{\rm s}}{0.19\:{\rm km}\:{\rm s}^{-1}} \right) M_{\odot}\:{\rm pc}^{-1}
\end{eqnarray}
for the isothermal case.

Multiple field strengths were explored, sampling values previously
used in the out-of-plane cases to keep the total magnetic pressure
component consistent. We apply $|B| = 10, 40$, and $65\:{\rm \mu G}$
and analyze the effects on isolated and colliding cases. These models
correspond to runs 2.x in Table~\ref{tab:all_runs}.

\subsubsection{Isolated cloud with embedded clump}

With uniform $B_{x}$ and $B_{y}$ fields, the GMC and clump collapse
along the direction of the field to form dense sheets perpendicular to
the field lines.  The timescales associated with their contraction are
of the order of the spherical free-fall time $t_{\rm ff}$, i.e.,
$\simeq 1.6$~Myr for the clump and 4.4~Myr for the GMC.  After the
initial collapse parallel to the magnetic field, the gas starts to
contract perpendicularly. Complete collapse of the clump takes much
longer as the gas motions are perpendicular to the magnetic field.

The isolated case is most similar to the models of \citet{Tomisaka_2014}. 
However, the embedded overdense clump dominates the gravitational 
collapse of the cloud. 
The line mass of the clump is 3450 $M_{\odot}\:{\rm pc}^{-1}$. 
Equation~\ref{eq:tomisaka} yields $\lambda_{\rm max} \approx$ 1660 
$M_{\odot}\:{\rm pc}^{-1}$ for 65 ${\rm \mu G}$, and even smaller 
values for 10 and 40 ${\rm \mu G}$. Thus, the maximum supported line mass
by in-plane magnetic fields is exceeded, and our simulations agree with
these results. A field strength of $\sim 135\:{\rm \mu G}$ could be used 
to support the clump, but this case was not explored.

\subsubsection{Colliding clouds}

For colliding clouds we again adopt a fiducial relative velocity of
$10~{\rm km\ s^{-1}}$ and study two different in-plane magnetic field
directions, i.e., parallel (i.e., $B_x$) and perpendicular (i.e.,
$B_y$) to the converging flow.  Here we set the collision time at
$t=0~{\rm Myr}$ as there is no more stable state to be reached.
Further, we only study a single collision speed as the dynamics are
dominated by the gravitational collapse of the clouds and clump.
Similar to the isolated model, the line mass of the clump and clouds
exceeds the maximum supportable by thermal and magnetic pressures. The
clouds collapse into flattened sheets perpendicular to the magnetic
field lines.
The timescales of gravitational collapse are again of the order of the
free-fall time.  In these highly collapsed scenarios, the clump is no
longer well tracked at late times due to numerical effects.

\subsection{Mixed field geometries}

While the previous sections describe two extremes, i.e., either the
cloud is maximally supported by magnetic fields (out-of-plane magnetic
field) or minimally (in-plane magnetic field), we now investigate a
combination of the two geometries. It represents a more realistic
situation as expected in 3D, where the magnetic field provides some
support against gravitational collapse, but cannot halt it completely,
if the cloud is supercritical.

\begin{figure*}[h]
\begin{center}
\includegraphics[width=2.0\columnwidth]{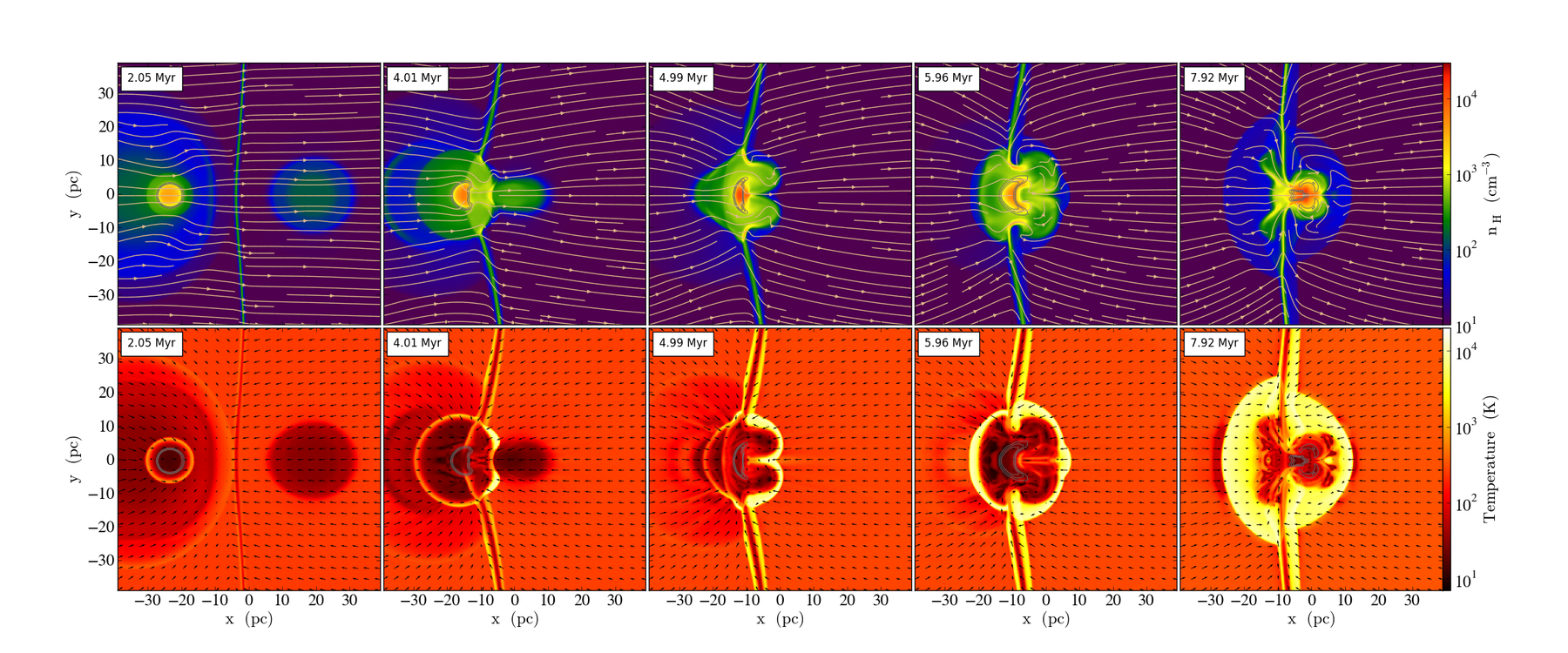}
\caption{\label{fig:map-Bmix_x_col}
Time evolution of colliding GMCs at 2.05, 4.01, 4.99, 5.96, and
7.92~Myr for the x-directed mixed field geometry (see run 3.C.1.0 in
Table~\ref{tab:all_runs}.)  Here, the total $B$-field magnitude is
near critical while an additional in-plane uniform field of
$B_{x}=10\:{\rm \mu G}$ is applied throughout the simulation.  ({\it
  top row}) Maps of $n_{\rm H}$ with magnetic fields represented by
streamlines and ({\it bottom row}) maps of temperature with black
vectors representing velocity are shown. The advective scalar defining
the clump is shown by grey contour lines, representing the scalar
value $S=0.25$, 0.5, and 0.75.  }
\end{center}
\end{figure*}

\begin{figure*}[h]
\begin{center}
\includegraphics[width=2.0\columnwidth]{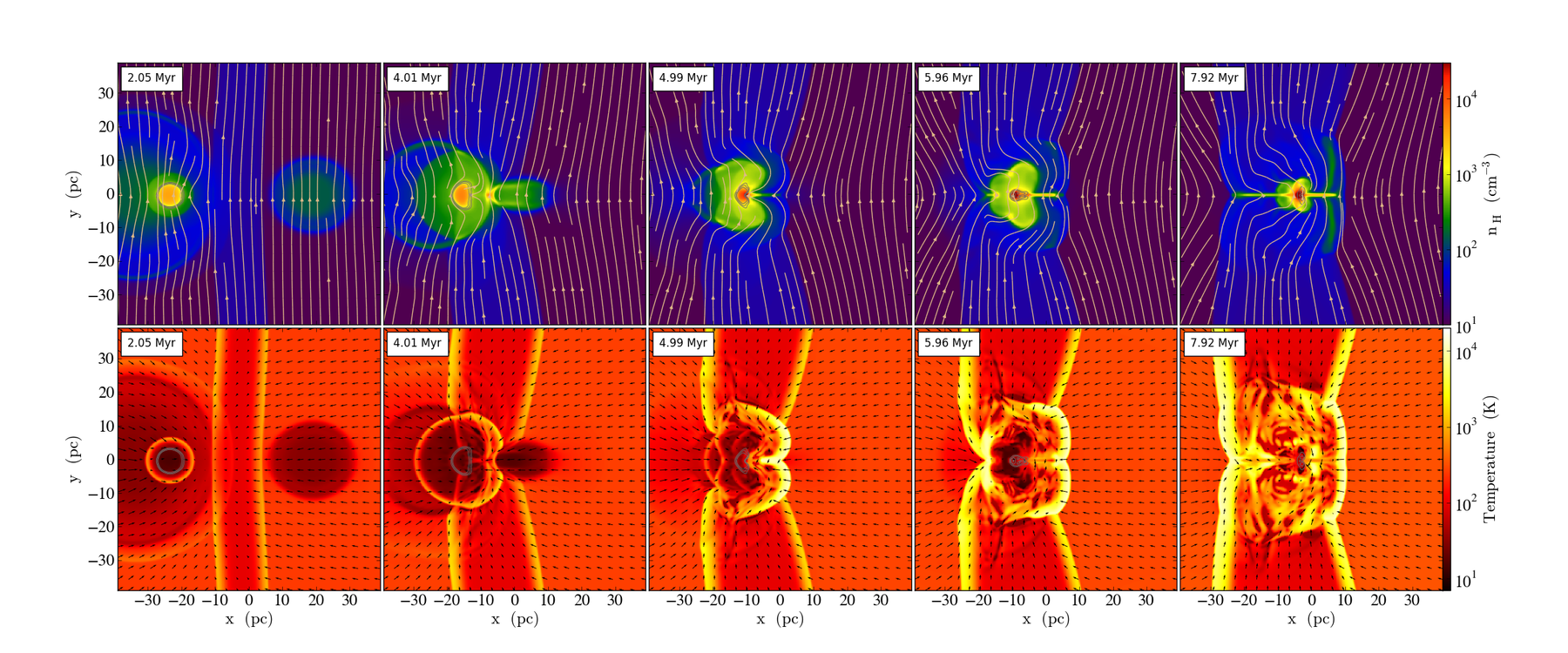}
\caption{\label{fig:map-Bmix_y_col}
Time evolution of colliding GMCs at 2.05, 4.01, 4.99, 5.96, and
7.92~Myr for the y-directed mixed field geometry (see run 3.C.2.0 in
Table~\ref{tab:all_runs}.)  Here, the total $B$-field magnitude is
near critical while an additional in-plane uniform field of
$B_{y}=10\:{\rm \mu G}$ is applied throughout the simulation.  ({\it
  top row}) Maps of $n_{\rm H}$ with magnetic fields represented by
streamlines and ({\it bottom row}) maps of temperature with black
vectors representing velocity are shown.  The advective scalar
defining the clump is shown by grey contour lines, representing the
scalar value $S=0.25$, 0.5, and 0.75.  }
\end{center}
\end{figure*}

In these cases, we assume a uniform in-plane magnetic field strength
of $10~{\rm \mu G}$ (along the $x$-axis
(Fig.~\ref{fig:map-Bmix_x_col}) or the $y$-axis
(Fig.~\ref{fig:map-Bmix_y_col})). The out-of-plane components are
chosen such that the total field strength has a magnitude equal to its
critical field strength (see Table~\ref{tab:GMC-properties} and runs
3.x in Table~\ref{tab:all_runs}).  The external medium has zero
out-of-plane magnetic field component, preserving the total field
strength of $|B_{0}| = 10~{\rm \mu G}$.  Such a field is both
density-dependent (as observed by \citet{Crutcher_2012}) and is
divergence-free.

\begin{figure}[h]
\begin{center}
\includegraphics[width=1.0\columnwidth]{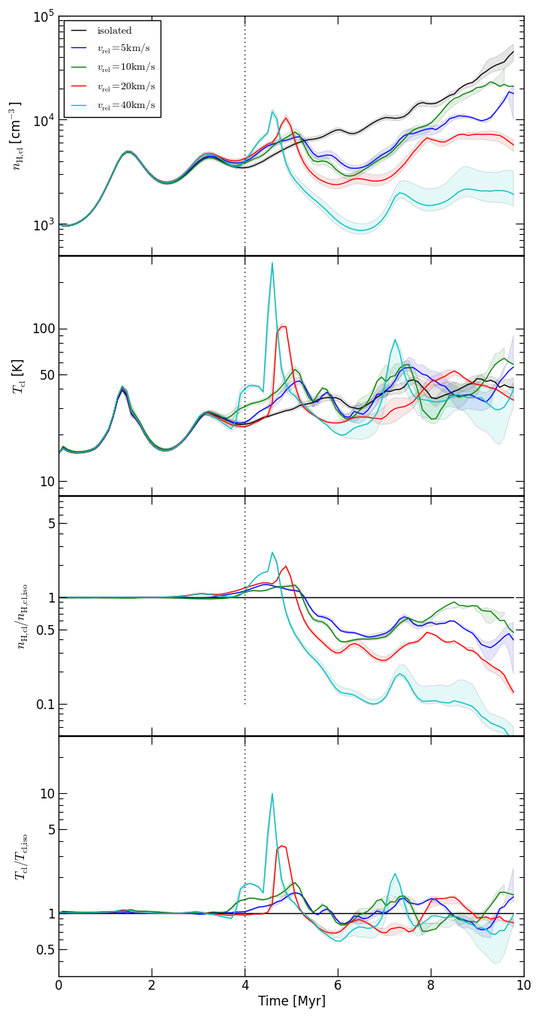}
\caption{\label{fig:summary-Bmix_x_col}
Average clump density ({\it Top panel}) and temperature ({\it 2nd panel}) over 
time comparing the effects of collision velocity for the 
$B_{\rm mix}$ field geometry case (see runs 3.C.1.x).  
Here, out-of-plane magnetic fields are near critical
($B_{\rm cl}=65\:{\rm \mu G}$, $B_{1}=40\:{\rm \mu G}$, and $B_{0}=10\:{\rm \mu G}$)
while an additional in-plane uniform field of $B_{x}=10\:{\rm \mu G}$
is applied throughout the simulation.
Collision velocities $v_{\rm rel}= 5,
10, 20, 40 {\rm km\ s^{-1}}$ are shown, along with the evolution of
the isolated GMC with clump.  
The ratios of the colliding cases compared to the isolated case for average 
density ({\it 3rd panel}) and temperature ({\it bottom panel}) are also shown. }
\end{center}
\end{figure}

\begin{figure}[h]
\begin{center}
\includegraphics[width=1.0\columnwidth]{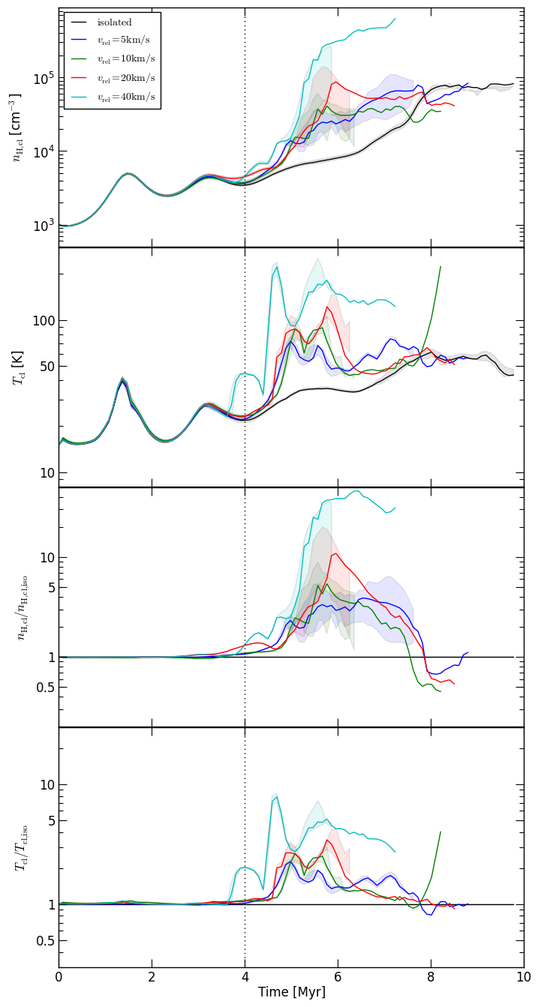}
\caption{\label{fig:summary-Bmix_y_col}
Average clump density ({\it Top panel}) and temperature ({\it 2nd panel}) over 
time comparing the effects of collision velocity for the
$B_{\rm mix}$ field geometry case (see runs 3.C.2.x).
Here, out-of-plane magnetic fields are near critical
($B_{\rm cl}=65\:{\rm \mu G}$, $B_{1}=40\:{\rm \mu G}$, and $B_{0}=10\:{\rm \mu G}$)
while an additional in-plane uniform field of $B_{y}=10\:{\rm \mu G}$
is applied throughout the simulation.
Collision velocities $v_{\rm rel}= 5,
10, 20, 40 {\rm km\ s^{-1}}$ are shown, along with the evolution of
the isolated GMC with clump.  
The ratios of the colliding cases compared to the isolated case for average 
density ({\it 3rd panel}) and temperature ({\it bottom panel}) are also shown. }
\end{center}
\end{figure}

\subsubsection{Isolated cloud with embedded clump}

As the out-of-plane magnetic field is near-critical and strong enough
to stabilize the GMC and clump, the early evolution is similar to the
out-of-plane case (see Figure~\ref{fig:summary-Bz_col}). A density
(and magnetic) gradient is quickly established to form an equilibrium
structure. However, gas also flows along the in-plane magnetic field.
Then the line mass of the cloud increases while the magnetic flux
remains constant. The clump and GMC gradually contract, although the
associated timescale is much longer than the free-fall time.  For a
larger ratio of in-plane to out-of-plane magnetic field, the evolution
is faster as the out-of-plane magnetic field is less dominant.

\vspace{10mm}

\subsubsection{Colliding clouds: head-on collisions}

We perform simulations of clouds colliding in this mixed-field
geometry, again investigating the effect of collision speed ($v_{\rm
  rel} = 5, 10, 20, {\rm and}~40~{\rm km\ s^{-1}}$).  Additionally,
the direction of the in-plane component of the magnetic field is
varied with respect to collision velocity (see runs 3.C.1.x and
3.C.2.x in Tab.~\ref{tab:all_runs}.)

Similar to the mixed-field isolated cloud case, the colliding clouds 
threaded by a mix of out-of-plane and  in-plane magnetic fields 
experience a larger compression compared to the purely out-of-plane 
clouds.  However, the direction of the uniform component of the magnetic 
field affects the late-time behavior of the clump.  

For partial fields parallel to the collision velocity (i.e.,
$x$-direction), there are temporary density increases of a factor of
$\sim 2-3$ during the collision, but the average clump density
actually decreases slightly at $\sim 5$~Myr and beyond, relative to
the isolated case (see Figure~\ref{fig:summary-Bmix_x_col}). This
rebound effect is greater for higher velocities. The shocks initially
compress the clump, then subsequently distort it, forming a
sickle-like shape. From Figure~\ref{fig:map-Bmix_x_col}, we see that
the original clump is broken apart due to the collision. The
temperature of the clump material is affected more significantly as
high velocity shocks dominate the mean clump temperature, temporarily
raising it to $\sim$few hundred K. The material cools to $\sim$tens of
K in the aftermath of the collision.

For partial fields perpendicular to the collision velocity (i.e.,
$y$-direction), the behavior is nearly identical for pre-collision
times $t<4~{\rm Myr}$. However, the different magnetic field geometry
causes the clump to be compressed in a different manner (see
Figures~\ref{fig:map-Bmix_y_col} and \ref{fig:summary-Bmix_y_col}).
In this case, the collision induces no sickle-shaped structure, but
rather the clump stays relatively compact, with the average density
increasing, but not rebounding. The material in the collisional flow
interface region freely falls into the overdense remnants of the cloud
and clump due to the orientation of the B-field.  Shocks are
continually created as the global flow and infalling material
interact, regulated by the magnetic fields. Late time behavior after
the collision reveals continuously increasing clump densities due to
infall, with elevated ($T\sim 50-100$ K) but roughly level
temperatures.

\subsubsection{Colliding clouds: off-axis collisions}

Our final model is a cloud-cloud collision in the mixed-field geometry 
with an in-plane uniform field of $B_{x}=10\:{\rm \mu G}$. We have 
$v_{\rm rel} = 10~{\rm km\ s^{-1}}$ and additionally apply $b=0.5 R_{1}$ 
to GMC 2 (see 3.D.0 in Tab.~\ref{tab:all_runs}).

\begin{figure}[t]
\begin{center}
\includegraphics[width=1.0\columnwidth]{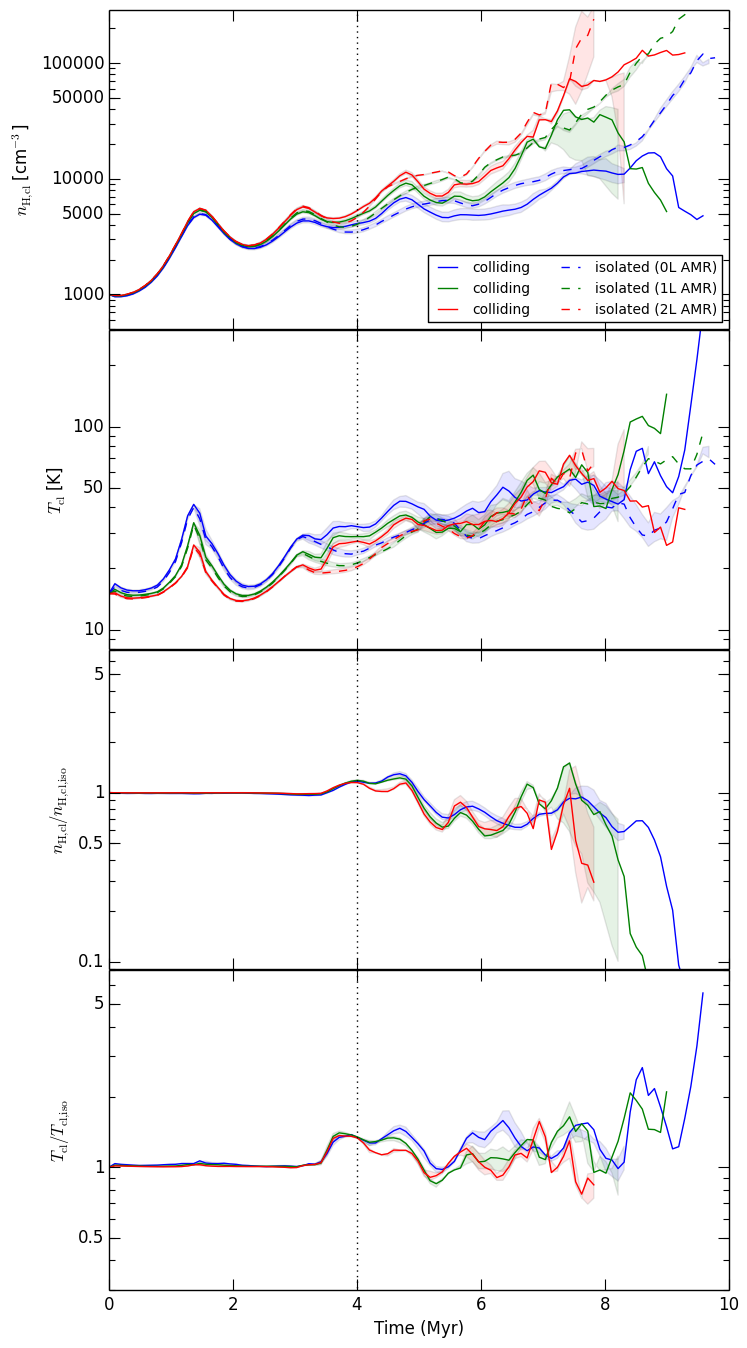}
\caption{\label{fig:resolution-Bmix_x_col_b}
A resolution study comparing time evolution of average clump density
({\it Top panel}) and temperature ({\it 2nd panel}) for the fiducial
case (see run 3.D.0). Models at the standard resolution ($1024^{2}$)
are compared with those run with one and two additional levels of AMR.
The ratios for average density ({\it 3rd panel}) and temperature ({\it
bottom panel}) compared to the isolated case at the particular
resolution, are also shown. }
\end{center}
\end{figure}

We designate this as our ``fiducial case'' and run the standard resolution,
along with one and two additional levels of AMR, giving a maximum
effective resolution of 0.0625~pc. We compare the effects of different
resolutions in Figure~\ref{fig:resolution-Bmix_x_col_b}.  Pre-collision 
densities are quite well converged, but begin to deviate as the shock waves 
and clump compression are realized at different resolutions.  Larger
initial differences are seen in the temperatures, where higher
resolutions lead to generally lower average clump temperatures. This is 
likely due to the initial shock created at the boundaries of the uniform 
clump as the density gradient is established. At higher resolutions, the
post-shock region contributes less to the overall clump material. 
Additionally, inspection of clump contours at the various resolutions 
revealed slightly different clump boundaries arising from the collision.
This could partially account for the greater discrepancies at later times.
While these resolution effects are not insignificant, the key results -- 
relative changes of a cloud collision with respect to the isolated case -- 
retain good agreement throughout the majority of the simulation (up to 
$\sim 8$~Myr).

\begin{figure*}[h]
\begin{center}
\includegraphics[width=2.0\columnwidth]{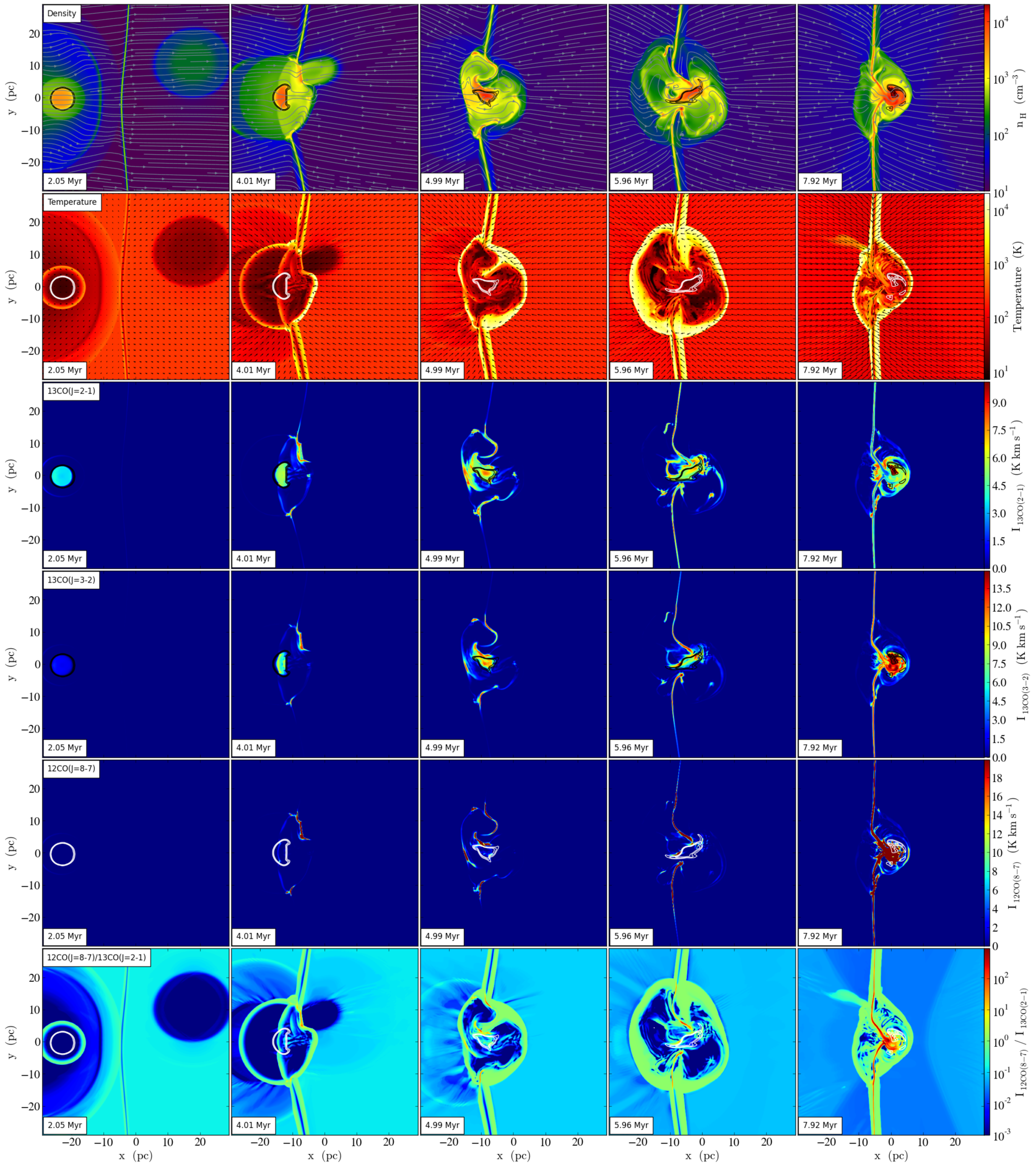}
\caption{\label{fig:map-Bmix_x_col_b}
Time evolution of colliding GMCs at 2.05, 4.01, 4.99, 5.96, and
7.92~Myr (1 level AMR version of run 3.D.0 in
Table~\ref{tab:all_runs}.)  Here, out-of-plane magnetic fields are
near critical ($B_{\rm cl}=65\:{\rm \mu G}$, $B_{1}=40\:{\rm \mu G}$,
and $B_{0}=10\:{\rm \mu G}$) while an additional in-plane uniform
field of $B_{x}=10\:{\rm \mu G}$ is applied throughout the
simulation. Furthermore, GMC 2 is offset at $b = 0.5 R_{1}$.  ({\it
  Row 1}): Maps of $n_{\rm H}$ with magnetic fields represented by
grey streamlines.  ({\it Row 2}): Maps of temperature with black
vectors representing velocity.  ({\it Row 3}): $^{13}$CO($J$=2-1) 
integrated intensity maps using PDR-based, temperature and density 
dependent volume emissivities.  ({\it Row 4}): Similarly derived 
$^{13}$CO($J$=3-2) line intensity maps.  ({\it Row 5}): Similarly 
derived $^{12}$CO($J$=8-7) line intensity maps.  ({\it Row 6}): 
$^{12}$CO($J$=8-7)/$^{13}$CO($J$=2-1) line intensity ratio maps.  
The advective scalar defining the clump is shown by black or white 
contour lines, representing the scalar value $S=0.25$, 0.5, and 0.75.
}
\end{center}
\end{figure*}

Figure~\ref{fig:map-Bmix_x_col_b} summarizes the entire fiducial run,
showing time evolution for maps of density, temperature, and common
observational bands of $^{13}$CO($J$=2-1), $^{13}$CO($J$=3-2), and 
$^{12}$CO($J$=8-7) as well as the $^{12}$CO($J$=8-7)/$^{13}$CO($J$=2-1) 
line ratio. These integrated intensity maps, based on outputs from the 
PDR modeling as potential observational diagnostics, are discussed in 
\S\ref{sec:diagnostics}.

Broadly speaking, the effect of a finite impact parameter for the GMC
collision results in a shearing velocity field and asymmetric
morphologies as various areas of the clump are compressed and
distorted.  GMC 2 can be seen contracting gravitationally as it
approaches the more massive GMC 1. Prior to the collision, parts of
GMC 1 and the clump are slightly compressed by the bounding shocks
arising from the colliding region of the ambient material. The
collision itself compresses parts of the clouds even further, as GMC 2
enters GMC 1 and impacts the clump from the north.  From the density
and temperature maps, shocks can be seen permeating the cloud material
and passing through the clump throughout the entire collision process.
At later times, the original clump material is distorted greatly and
even breaks apart into a few pieces, but the densest material remains
inside the main clump region.

\begin{figure*}[t]
\begin{center}
\includegraphics[width=2.0\columnwidth]{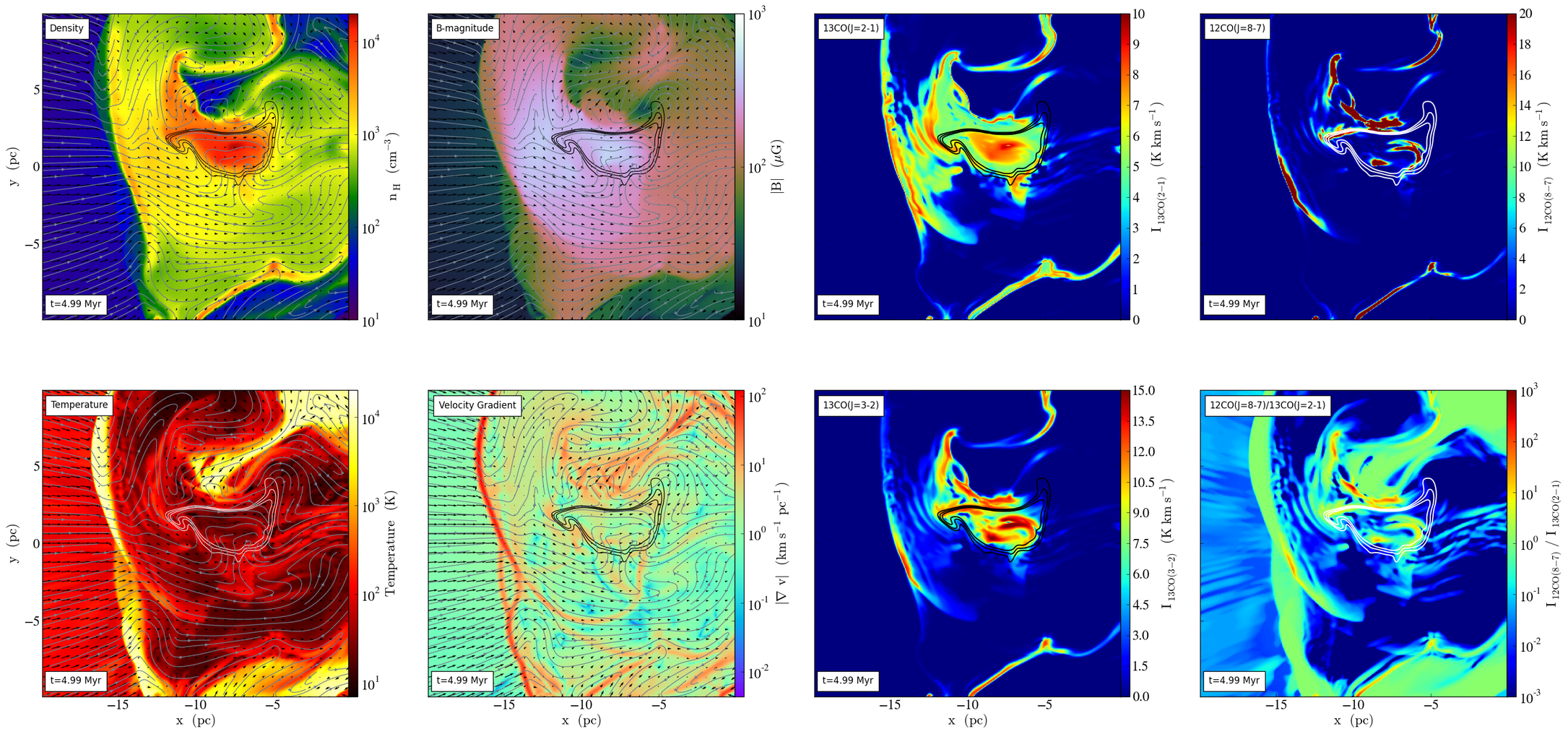}
\caption{\label{fig:map-Bmix_x_col_b_clump}
Zoomed in maps of the clump at $t=4.99~{\rm Myr}$, near the time of
maximum compression due to the collision.  This is the x-directed
mixed field geometry case with $b=0.5 R_{1}$ (2 level AMR version of
run 3.D.0 in Table~\ref{tab:all_runs}.)  Here, the total $B$-field
magnitude is near critical while an additional in-plane uniform field
of $B_{x}=10\:{\rm \mu G}$ is applied throughout the simulation.
({\it left 4 figures}) Maps of density ($n_{\rm H}$), temperature, $B$-field
magnitude, and velocity gradient magnitude are shown. Grey streamlines
indicate magnetic field structure while velocities are represented by
the black vectors.  ({\it right 4 figures}) Maps of $^{13}$CO($J$=2-1),
$^{13}$CO($J$=3-2), and $^{12}$CO($J$=8-7) intensity, as well as a map of
$^{12}$CO($J$=8-7)/$^{13}$CO($J$=2-1) line intensity ratio are shown.  The
advective scalar defining the clump is shown by black or white contour lines,
representing the scalar value $S=0.25$, 0.5, and 0.75.  }
\end{center}
\end{figure*}

Peak compression due to the collision occurs near 5 Myr (third column
in Figure~\ref{fig:map-Bmix_x_col_b}). We investigate this further by
zooming in on the clump at this timestep and mapping various key
quantities.  This is shown in
Figure~\ref{fig:map-Bmix_x_col_b_clump}. The density, temperature,
magnetic fields, and velocity gradient in the regions surrounding and
including the clump are analyzed.  Integrated intensity maps are also
shown in this figure and discussed in \S\ref{sec:diagnostics}.

The clump, initially a uniform cylinder, remains relatively distinct
and contiguous, though at this timestep it is undergoing compression
and distortion due to the cloud collision. What was once GMC 2 can be
seen as the denser (few $\times 10^{3}\:{\rm cm^{-3}}$) material that
has punched into GMC 1 and is impacting the clump from the north.  The
average clump density is $n_{\rm H} \sim 10^{4}\:{\rm cm^{-3}}$,
embedded in GMC material of $\sim 10^{2}$--$10^{3}\:{\rm cm^{-3}}$.

The clump temperature, on the other hand, is not particularly distinct
from the surrounding material, generally at a few 10s of K. Shocks of
a few 100s of K are seen propagating through the clump and cloud. The
high temperature material ($\sim 10^{4}\:{\rm K}$) due to the strong
shock created by the collision with GMC 2 has penetrated GMC 1, but
has not reached the clump.

The magnetic fields can be seen corresponding closely to the density
morphology of the GMC, with field strength generally increasing with
density. The B-fields have strengths of $\sim 100\:{\rm \mu G}$ in the
compressed GMC material and peak at a few hundred ${\rm \mu G}$ within
the clump and nearby regions. The initial in-plane fields, uniform and
directed along the collision axis, remain mostly uniform, except for
where the GMCs have been disrupted. Complex field structures arise
within the clump and cloud material. In this case, there is a loose
correlation between magnetic field direction and the direction of
infalling gas flow to the clump.

The velocity gradient map shows detailed structure of the many shocks
propagating throughout the cloud. The strongest gradients can be seen
corresponding with the shocked GMC-envelope interface, as well as the
GMC-GMC collision region.  The velocity magnitudes show some turbulent
motion being produced by the collision.

\begin{figure}[t]
\begin{center}
\includegraphics[width=1.0\columnwidth]{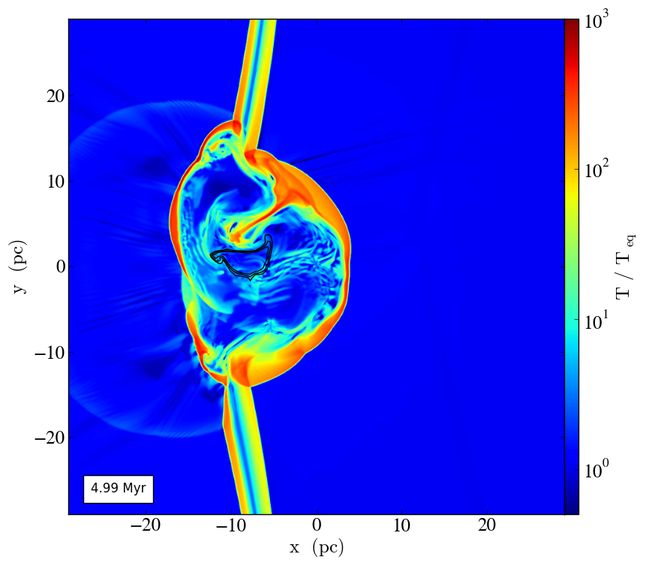}
\caption{\label{fig:map-noneq_temp}
A map of the ratio of actual simulation temperature to the
density-based equilibrium temperature at $t=4.99~{\rm Myr}$ for the
2 level AMR fiducial case.  The advective scalar at values $S=0.25$,
0.5, and 0.75 defining the clump is shown by black contour lines. }
\end{center}
\end{figure}

To illustrate the effects of our treatment of nonequilibrium heating
and cooling, Figure~\ref{fig:map-noneq_temp} compares the differences
in temperatures between the nonequilibrium cooling/heating functions
developed in this paper and a cooling/heating curve that assumes
equilibrium temperatures. Differences primarily occur in the shocked
regions, as material is shock heated out of equilibrium. The
temperature maps, upon which the observational diagnostics heavily
depend, would exhibit very different behavior had only a simple
equilibrium cooling/heating curve been used.

\section{Observational Diagnostics}
\label{sec:diagnostics}

Here we briefly outline two potential methods of observationally
diagnosing GMC collisions, based on emission of high-$J$ CO
lines. However, given the idealized 2D nature of the simulations
presented so far, we defer more detailed discussion to a future paper
that will consider the outputs from 3D simulations.

\subsection{Integrated Intensity Maps}

Using the outputs from our PDR modeling (method described in
\S\ref{sec:heatcool-diagnostics}), we create CO integrated intensity 
maps from the simulation outputs. Note that the local emissivity of 
CO lines does take into account the optical depth of an associated 
PDR layer, of given total column density that depends on local volume
density. To illustrate the method, we perform post-processing on the 
colliding case ($v_{\rm rel}=10\:{\rm km/s}$) with an impact 
parameter of $b=0.5 R_{1}$ and $B_{x}$-oriented mixed fields, our 
fiducial model.

The diagnostics portions of Figures~\ref{fig:map-Bmix_x_col_b} and
~\ref{fig:map-Bmix_x_col_b_clump} show integrated intensity maps of
common observational bands of $^{13}$CO($J$=2-1 and 3-2) and
$^{12}$CO($J$=8-7) as well as the $^{12}$CO($J$=8-7)/$^{13}$CO($J$=2-1) line
ratio. These maps assume a depth of 1~pc in the $z$ direction and a
cloud distance of 3~kpc. Note also that the adopted abundance ratio of
$^{13}$C to $^{12}$C is 1/60.
We see that the CO emission lines trace molecular gas in general, with
the higher-$J$ lines indeed probing more strongly shocked regions.  As
$J$ increases, higher temperature material is traced, with shock
fronts of varying strengths being followed. This occurs even for low
values of $n_{\rm H}$.  While these line emissivities are most
strongly affected by temperature, they are also tracers of high
density due to the higher critical density of the high-$J$ transitions
and the dependence of $n_{\rm CO}$ on $n_{\rm H}$. Thus, lower
temperature, high $n_{\rm H}$ gas is also revealed.

$^{13}$CO($J$=2-1) and $^{13}$CO($J$=3-2) intensity maps show fairly
similar structures, primarily tracing high-density material as well as
higher temperature regions. The $^{12}$CO($J$=8-7) map, however, accentuates
more strongly shocked regions, closely tracing the high-temperature
dense regions.

Strongly shocked, high temperature, high density gas -- potentially a
signature of cloud-cloud collisions -- produces the strongest intensity
of higher-level lines.  Emissivities at certain $J$ levels as well as
their ratios can act as diagnostics of a wide range of conditions and
potentially determine shock properties and physical conditions in the
affected gas.

The final line ratio map further traces high-temperature, high-density
material, and de-emphasizes low-temperature, high-density material.
The $^{12}$CO($J$=8-7)/$^{13}$CO($J$=2-1) line ratio could be an efficient
tracer of cloud collisions.

\begin{figure}[h]
\begin{center}
\includegraphics[width=1.0\columnwidth]{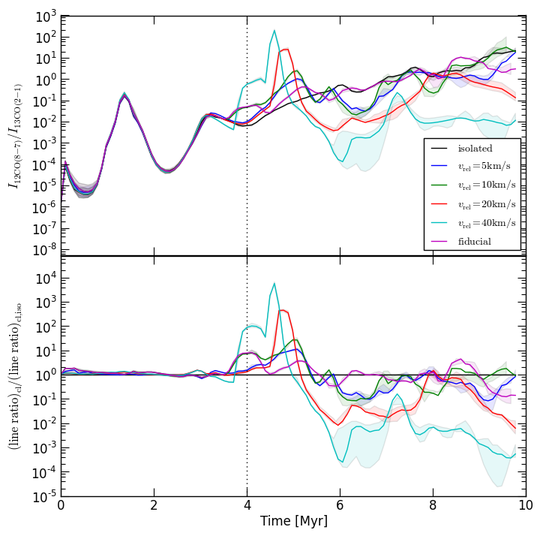}
\caption{\label{fig:emiss-Bmix_x_col_b_clump}
Average $^{12}$CO($J$=8-7)/$^{13}$CO($J$=2-1) line intensity ratio ({\it top})
over time comparing the effects of collision velocity for the $B_{\rm
  mix}$ field geometry case (see runs 3.C.1.x).  Here, out-of-plane
magnetic fields are near critical ($B_{\rm cl}=65\:{\rm \mu G}$,
$B_{1}=40\:{\rm \mu G}$, and $B_{0}=10\:{\rm \mu G}$) while an
additional in-plane uniform field of $B_{x}=10\:{\rm \mu G}$ is
applied throughout the simulation.  Collision velocities $v_{\rm rel}=
5, 10, 20, 40 {\rm km\ s^{-1}}$ are shown, along with the evolution of
the isolated GMC with clump.  
The ratios of the colliding cases compared to the isolated case ({\it
bottom}) are also shown. }
\end{center}
\end{figure}

Figure~\ref{fig:emiss-Bmix_x_col_b_clump} explores this potential
cloud-collision signature. The average $^{12}$CO($J$=8-7)/$^{13}$CO($J$=2-1)
line intensity ratio within the clump is calculated and followed over
time for a set of isolated and colliding cases.  From these results,
we see that this parameter is an excellent tracer of cloud
collisions. While the clump in the isolated case (once it settles into
a relatively stable state) retains a value of this intensity ratio of
$\sim 1-10$, a clump experiencing a GMC collision sees much larger
values of the line ratio, even reaching $> 10^{3}$ for $v_{\rm rel}=
40~{\rm km\ s^{-1}}$.  Collision velocities as low as $v_{\rm rel}=
5~{\rm km\ s^{-1}}$ show an excess of a factor of $\sim 10$ with
respect to the isolated case.

\subsection{Spectra}

\begin{figure}[h]
\begin{center}
\includegraphics[width=1.0\columnwidth]{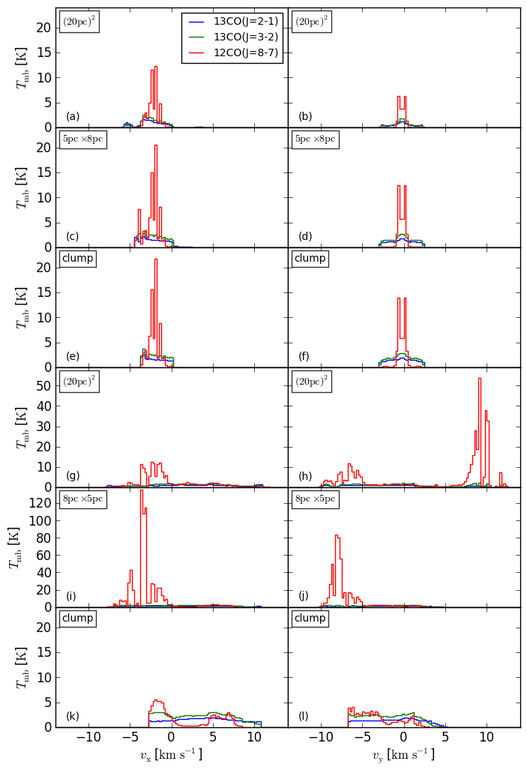}
\caption{\label{fig:spectra}
Synthetic spectra for $^{13}$CO($J$=2-1), $^{13}$CO($J$=3-2), and
$^{12}$CO($J$=8-7) are shown for ({\it top six: (a) - (f)}) the isolated 
case and ({\it bottom six: (g) - (l)}) colliding case, at $t=4.99~{\rm Myr}$.
The subplots denote emission analyzed from different volumes ($x \times y$ 
values listed below, and assuming 1~pc extent in the $z$ direction) in the
simulation: a $20\:{\rm pc}\times20\:{\rm pc}$ box centered on the clump, a
smaller $5\:{\rm pc}\times8\:{\rm pc}$ (isolated case) or  
$8\:{\rm pc}\times5\:{\rm pc}$ (colliding case) region containing the clump,
and contribution solely from the clump material, defined by the scalar value 
$S>0.5$. The left column shows spectra derived from $v_{x}$ (i.e., a view 
along the collision axis), while the right column shows spectra derived from 
$v_{y}$ (i.e., a view perpendicular to the collision axis).
Velocity bins of $0.25\:{\rm km}\:{\rm s}^{-1}$ are used. }
\end{center}
\end{figure}

From the simulations, synthetic spectra were created in order to
provide a more direct comparison with observed cloud kinematics. While
the initial conditions and 2D geometry are fairly idealized, we expect
these diagnostic methods to be of general use, e.g., once outputs from
3D simulations are available.  Emission line spectra of various
observational volumes within the simulation box for the isolated and
colliding fiducial case are shown in Figure~\ref{fig:spectra}.

The isolated case shows a narrow velocity range of dense gas tracers
There is also a relatively strong peak in $^{12}$CO($J$=8-7) as the
cloud pinches in on itself due to the presence of in-plane magnetic
fields, but these shocks occur at low velocities. The chosen volume has
little effect on the spectra as the main features are localized around 
the clump region. The same features are present in both lines of sight, 
with greater asymmetry in the x-velocity simply due to the off-center 
initial position of the clump relative to GMC 1.
The spectra for the colliding case show a much wider velocity spread in
each of the CO emission lines. In the $20\:{\rm pc}\times20\:{\rm pc}$ 
box, the emission peaks in $^{12}$CO($J$=8-7) at multiple narrow 
velocity bands correspond to the strongest shocks as seen in 
Fig.~\ref{fig:map-Bmix_x_col_b_clump}. The high emissivity feature in 
$^{12}$CO($J$=8-7) for the $y$-velocity indicates strong shocks traveling
northward around the collision region. The strong features present in the 
8~pc by 5~pc region but not the clump material represent shocks 
compressing, but not yet propagating through, the clump. These shocks,
directed in the negative-$x$ and $y$ directions, indicate the collision 
with GMC 2. 

\begin{figure}[t]
\begin{center}
\includegraphics[width=1.0\columnwidth]{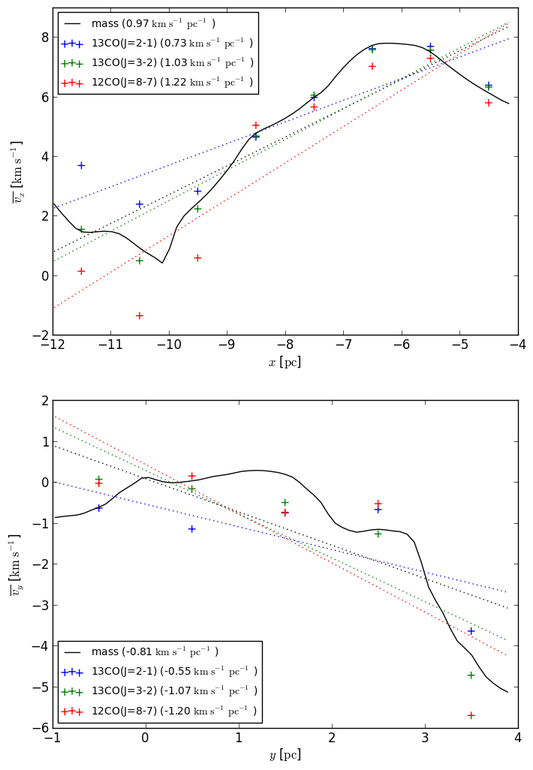}
\caption{\label{fig:vel-grad}
Mean velocities of an 8~pc by 5~pc rectangular region around the clump
at $t=4.99~{\rm Myr}$, along the ({\it top}) $x$-direction and ({\it bottom})
$y$-direction.
Black lines represent line of sight velocities weighted by the mass 
distribution of the region. 
Blue crosses are intensity-weighted mean velocities derived from 
$^{13}$CO($J$=2-1) spectra of 1~pc wide strips that evenly divide the region
along the line of sight.
Green and red crosses denote similarly calculated mean velocities from
$^{13}$CO($J$=3-2) and $^{12}$CO($J$=8-7), respectively.
The colored dotted lines show the best linear fit to each corresponding set of 
points, with the value of this gradient displayed in parenthesis in the legend.
}
\end{center}
\end{figure}

Next we measure line-of-sight velocities and velocity gradients in the
8~pc by 5~pc rectangular region around the clump in the fiducial simulation
at the time of maximum compression using the $^{13}$CO($J$=2-1) spectra.
We compare gradients derived from the total mass distribution and 
those derived from the intensities of $^{13}$CO($J$=2-1), $^{13}$CO($J$=3-2) 
and $^{12}$CO($J$=8-7) spectra.
Figure~\ref{fig:vel-grad} shows the mean velocities and derived velocity
gradients of the clump material along orthogonal lines of sight. The
mean gradients are 0.97 and -0.81 ${\rm km\ s^{-1}\:pc}^{-1}$ for
spectra measured along lines of sight perpendicular and parallel
to the collision direction, respectively. Somewhat smaller gradients are 
derived from lower $J$ CO lines, and larger gradients from higher $J$ lines.

Velocity gradients have been measured observationally within IRDCs and
GMCs. For example, \citet{Ragan_ea_2012} found velocity gradients of
2.4 and 2.1 ${\rm km\ s^{-1}\ pc^{-1}}$ within sub-pc regions of IRDCs
G5.85-0.23 and G24.05-0.22, based on observations of NH$_3$ (1, 1).
\citet{Henshaw_ea_2014} found values of 0.08, 0.07, and 0.30 ${\rm
  km\ s^{-1}\ pc^{-1}}$ on $\sim2\:{\rm pc^{-1}}$ scales and larger
local gradients (1.5 − 2.5 ${\rm km\ s^{-1}\ pc^{-1}}$) on sub-parsec
scales in IRDC G035.39−00.33, based on centroid velocities of the
dense gas tracer N$_2$H$^+$(1-0). \citet{Hernandez_Tan_2015arXiv} derived a
mean velocity gradient of 0.24~${\rm km\ s^{-1}\ pc^{-1}}$ of 10 IRDC
clumps, based on $^{13}$CO(1-0) emission.

The gradients seen in our simulated clump are somewhat larger than those
observed towards IRDCs by \citet{Hernandez_Tan_2015arXiv}, which may 
indicate these IRDCs are not being disturbed kinematically by the kind 
of collision modeled in our fiducial simulation. However, the results of 
a range of 3D simulations and a wider variety of viewing angles are needed 
before more definitive conclusions can be drawn from such comparisons.

\section{Discussion and Conclusions}
\label{sec:conclusion}

We have explored a wide range of parameter space of magnetized GMC-GMC
collisions.  We performed idealized 2D simulations from the GMC-scale
down to $\sim 0.1$ parsec scales, allowing us to study in detail the
structure of GMCs undergoing collisions. In particular, we focused on
a clump embedded in a GMC, aiming to isolate the effects of various
parameters on the evolution of clump material. 

We began by developing new heating and cooling functions that depend
on density, temperature and extinction, based on the method of
VLBT2013. We combined the results from the PDR codes \textsc{PyPDR}
and \textsc{Cloudy} to create arrays of heating and cooling rates that
span the atomic to molecular transition, allowing treatment of a
multi-phase ISM and including the thermal instability of warm and cold
atomic media. Our heating and cooling functions return self-consistent
rates from densities ranging from $n_{\rm H} = 10^{-3}-10^{6}~{\rm
  cm^{-3}}$ and temperatures ranging from $T = 2.7-10^{5}~{\rm
  K}$. This enabled us to study non-equilibrium temperature conditions
that are present in the shocked material of colliding clouds. Further,
we similarly derived emissivity arrays for various common
observational bands of CO, allowing us to simulate synthetic
observations via post-processing.  

In terms of MHD simulations, our models tracked an initially
magnetically subcritical clump embedded within a GMC. We investigated
different collision velocities, impact parameters, magnetic field
strengths and orientations and their effects on colliding versus
isolated GMCs. For the maximally supportive out-of-plane $B$-field
cases, we reported GMC collisions at typical velocities causing
density increases of a factor of $\sim 2-3$, with a rebound resulting
in relatively lower average densisties post-collision. During the
collision, average clump temperatures were increased by a factor of up
to $\sim 10-20$ due to shocks dominating the clump material before
settling back to $\sim 15-30$~K. Collisions with impact parameter
between the GMCs produced similar levels of contraction with less
exaggerated effects for higher impact parameters. However, these types
of collisions involve strongly shearing velocity fields that produced
asymmetries and more filamentary structure, as well as imparting
angular momentum to the resulting cloud.

Mixed-field geometries resulted in relative increases of density and
temperature at levels similar to the out-of-plane case. However,
late-time behavior of the clouds showed eventual contraction, as
material is able to flow along the field lines and slowly accumulate
onto the clump.

Analysis of CO line emissivities provided a way to track shocks of
various strengths. In particular, the average value of the
$^{12}$CO($J$=8-7)/$^{13}$CO($J$=2-1) ratio within a clump that was
undergoing a collision versus one in an isolated cloud resulted in
differences of a factor of up to $\sim 10^{4}$ for typical collision
velocities. Even slow collisions of $v_{\rm rel} = 5~{\rm km\ s^{-1}}$
showed excesses of a factor of $\sim 10$ in this parameter. This may
be a useful diagnostic signature of cloud collisions. Spectra and velocity
gradients of molecular line emission around dense gas clumps may also
provide tests of cloud collisions as a triggering mechanism for their
formation.

One caveat of the presented models is that all the shocks are in the
context of ideal MHD. Also, the effects of initial GMC turbulence,
ambipolar diffusion, star formation, and stellar feedback have not
been addressed in this study, but are planned in future 3D
models. However, star formation and stellar feedback are not likely to
be too important in comparing simulation outputs with some ISM clouds,
such as Infrared Dark Clouds. A more complete study of observational
diagnostics with comparison to cases in the Galaxy is planned in a
subsequent paper that will analyze 3D simulations and include initial
GMC turbulence.

\acknowledgments We thank Fumitaka Nakamura and Shuo Kong for useful
discussions. J.C.T. acknowledges NASA Astrophysics Theory and
Fundamental Physics grant ATP09-0094.

\bibliographystyle{apj}

\clearpage

\clearpage

\end{document}